# Quantifying Atomic-Scale Transition Rates Using Stochastic Resonance Spectroscopy

Nicolaj Betz[1,2], Gregory McMurtrie[1], Max Hänze[1,3], Vivek Krishnakumar Rajathilakam[1],

Laëtitia Farinacci[1,4], Susan N. Coppersmith[5], Susanne Baumann[1] and Sebastian Loth[1,2,3]

[1]University of Stuttgart, Institute for Functional Matter and Quantum Technologies, 70569 Stuttgart, Germany.
[2]Center for Integrated Quantum Science and Technology (IQST), University of Stuttgart, 70569 Stuttgart, Germany.
[3]Max Planck Institute for Solid State Research, 70569 Stuttgart, Germany.
[4]Carl-Zeiss-Stiftung Center for Quantum Photonics Jena – Stuttgart – Ulm, Germany.
[5]School of Physics, University of New South Wales, Sydney, Australia.

## ABSTRACT:

A system's internal dynamics and its interaction with the environment can be determined by tracking how external perturbations affect its transition rates between states. Quantitative measurements of these rates are crucial for optimizing quantum systems at the atomic scale but are challenging, as these dynamics are often faster than experimental observation capabilities. Here, we show that driving a stochastic system periodically enables quantitative determination of its transition rates over a wide frequency range spanning from kilohertz to gigahertz. To perform this quantitative extraction, we provide an analytical model that is applicable irrespective of the details of the studied system. We name this method stochastic resonance spectroscopy (SRS). We apply it to quantify spin switching in individual magnetic atoms, Néel state reversal in nano-antiferromagnets and quasiparticle transport through Yu-Shiba-Rusinov states. We anticipate that the ability to characterize broadband stochastic dynamics at the atomic scale will enable insights into an even wider range of systems, e.g. dynamics driven by light-matter interaction. Another exciting prospect is the investigation of systems that exhibit non-Markovian dynamics due to correlations with the environment.

## I. INTRODUCTION

Stochastic dynamics are ubiquitous in nature and crucial for understanding quantum systems at the atomic scale. A system's stochastic dynamics reflects the energy landscape of its states, the transition probabilities between them, and the system's interaction with the environment. Full characterization of these properties is typically accomplished through recording state trajectories that show the stochastic switching in real time [1, 2], which requires experimentally accessible observables to be recorded faster than the fastest rate of the system under investigation. Yet, detecting fast transition rates in nanoscale structures presents significant challenges arising from both instrumental [3, 4] and fundamental limitations [5].

The scanning tunneling microscope (STM) is an ideal tool to measure such nanostructures on the atomic-scale, but on these short length scales, the challenges are particularly pronounced. However, in special cases, the interaction between the investigated system and its environment stabilizes the dynamics which makes them observable in real time: for example, magnetization or charge switching have been observed in large nanostructures [6, 7, 8, 9, 10, 11]. With sufficient stabilization from the environment, even dynamics of single atoms can be resolved [12, 13, 14]. This recently enabled the observation of an interesting effect called stochastic resonance in the slow spin switching of Fe atoms on a thin insulator with fourfold symmetry [15], the structural switching of atoms [16] and molecules [17] on surfaces, and in the dynamics of single-electron levels of a quantum-dot [18]. Stochastic resonance is an intriguing phenomenon in which the switching dynamics synchronizes with an external periodic drive when the transition rates match the frequency of the drive [19, 20, 21]. This phenomenon is universal and has been observed in a vast variety of systems – from the periodicity of ice ages [22] to the dynamics of electronic devices [23] – and plays a critical role in complex structures such as neurons [24]. The exploitation of this phenomenon at the atomic scale is a promising route to gain deeper understanding of the dynamics of quantum systems.

On the atomic scale, the transition rates of individual quantum systems on surfaces are in most cases too fast to be tracked in real time [25, 26]. Relaxometry methods such as pump-probe spectroscopy [27, 28, 29, 30], spin-echo measurements in electron paramagnetic resonance [31, 32], or time-resolved luminescence [33, 34] typically resolve relaxation dynamics of the undriven systems. By contrast, the driven dynamics remain mostly unexplored. These are however crucial for many technological advancements, e.g. in spintronics, where magnetic switching occurs within picoseconds [35], or quantum computing, where qubit manipulation requires precise control at nanosecond speeds [36].

Here, we demonstrate that the phenomenon of stochastic resonance can be harnessed for quantitative measurements of transition rates of free and driven dynamics. We term this broadband method *stochastic resonance spectroscopy (SRS)*. It is applicable to a variety of discrete systems with transition rates ranging over six orders of magnitude in frequency. First, we show that SRS is not limited to spins but applicable to other dynamics by performing the first direct measurement of a Yu-Shiba-Rusinov state lifetime [37, 38, 39]. Second, we develop the theoretical framework of the method which allows for the quantitative extraction of the transition rates and establishes that SRS requires only two conditions to be fulfilled: (i) the rates need to be modulated by a drive, and (ii) the measured signal needs to be sensitive to the state occupation. Starting from a generic scenario we show that this framework gives experimental insights into the energy dependence of transition rates. We extend our study to the case of a multi-state system, Fe on a Cu site on $Cu_2N$ [40], and address both the free and driven dynamics. This allows us to identify the specific states involved in the stochastic processes as a function of applied bias voltage.

## II. OVERVIEW OF STOCHASTIC RESONANCE SPECTROSCOPY

In stochastic resonance spectroscopy, we use a scanning tunneling microscope (STM) and apply an AC voltage drive, $V_{ac}$, at frequency $f$ and record the impact of varying the frequency of this electrical drive on the time-averaged tunnel current.

The resulting SRS signal as a function of drive frequency is shown in Fig.1 for three different systems: an individual $S = 2$ Fe atom on a Cu binding site on $Cu_2N/Cu(100)$ (green), a nano-antiferromagnet consisting of four Fe atoms assembled on the same substrate (light green), and a Yu-Shiba-Rusinov (YSR) state induced by Fe atoms on V(100) (dark green). Topographies of the different systems are shown in Fig. 1a-c. The measurements on the spin systems are performed using a spin-polarized tip while the measurement on the YSR state uses a superconducting tip.  For all three systems, the SRS signal appears as a single step that smoothly transitions between two levels over approximately one order of magnitude in frequency for logarithmically spaced frequencies. We find that the step has the same shape and width for all three systems but is centered around 31.7 kHz for the nano-antiferromagnet, 0.27 GHz for the individual Fe atom and 0.78 GHz for the YSR state. The following sections provide physical meaning to the position, width and amplitude of the signal and show how SRS measurements can be performed to gain insights into the transition rates of a broad range of systems.

The broad applicability of SRS stems from the fact that only two requirements need to be fulfilled: (i) it needs an external control parameter that can influence the transition rates of the investigated, discrete system and (ii) it must be possible to record the time average of an observable that depends on the state of the investigated system. Given these requirements, most SRS measurements can be described using a two-state approximation that results in the following formula for the measured current, $I$, as a function of frequency:

$$I(f) = I_{SR} \frac{1}{\Omega^2 + 1} + I_0 = I_{SR} \frac{\Gamma^2}{\Gamma^2 + (2\pi f)^2} + I_0 \,, \tag{1}$$

where the $I_{SR}$ represents the height of the observed step, $I_0$ is the offset current and $\Omega = 2\pi f / \Gamma$. Importantly, $\Gamma$ is the sum of the time-averaged excitation $\Gamma_+(t)$ and relaxation $\Gamma_-(t)$ rates, i.e. $\Gamma = \langle \Gamma_+(t) + \Gamma_-(t) \rangle$. On a logarithmic frequency scale, the Lorentzian function of eq. 1 appears as a step with fixed width centered around $\Omega = 1$, where the external drive frequency matches exactly to the total transition rate $\Gamma$ of the system. Hence, the position of the step is at $\Gamma/2\pi$ and gives a direct measure of the total transition rate of the system.

Fitting the experimental data in Fig. 1e with eq. 1 (solid lines) yields transition rates of $\Gamma_{Fe} = (1.68 \pm 0.05) \cdot 10^9/s$ for the single Fe atom, $\Gamma_{AFM} = (199 \pm 5) \cdot 10^3/s$ for the Néel state switching of the nano-antiferromagnet, and $\Gamma_{\mathrm{YSR}} = (4.91 \pm 0.06) \cdot 10^9/\mathrm{s}$ for the quasiparticle transport through the YSR state.

The Fe spin systems on $Cu_2N$ have been characterized previously by inelastic tunneling spectroscopy [40, 8]. Because of their large easy-axis magnetic anisotropy both behave as effective two-state systems at small bias voltage, which allows for quantitative comparison to our SRS measurements. We find that the difference in rates between the antiferromagnet and the individual Fe atom matches the difference of transition rates estimated from spin state lifetimes of similar structures reported in literature [41, 8]. In these experiments, we satisfy the requirements for SRS, by using a spin-polarized tip (meeting requirement (ii)) and modulating the bias voltage (achieving (i)). The measured switching rate determined with SRS for the spin of the single Fe atom is also consistent with the lifetime that we obtained by performing all-electronic pump probe measurements, which yields a spin relaxation time of $0.6 \pm 0.1$ ns corresponding to a relaxation rate of $\Gamma_{-,Fe} = (1.7 \pm 0.4) \cdot 10^9/s$ (Supplemental Material Fig. S1 [42]). Both results agree within their error bars, but SRS provides significantly tighter confidence intervals.

Importantly, these two SRS measurements demonstrate the large dynamic range of this technique and highlight its high accuracy: the switching rate of the individual Fe atom differs from the antiferromagnet's by four orders of magnitude, yet both rates are measured with a relative uncertainty of only 3 %. This is the case because logarithmic scaling sharpens the observed step and thus improves the accuracy of the measurements.

In contrast to spin dynamics, the quasiparticle lifetimes of YSR states and their transport properties have scarcely been studied, thus far. Their intrinsic relaxation times were measured indirectly by shot noise measurements or by fitting the nonlinear dependence of the tunnel current on the junction conductance [43, 44, 45]. SRS instead provides direct access to the YSR state lifetime. In addition, occupying a YSR state requires a minimal excitation energy, which makes meeting the requirements for SRS relatively simple. Requirement (i) is directly met by modulating the bias voltage around the excitation threshold. In addition, the tunnel current is directly sensitive to the occupation of the state [43], which satisfies requirement (ii) without the need for tip functionalization. The rate extracted from our SRS measurements agrees with an estimate obtained by tuning the tunnel junction conductance as in [43] (Supplemental Material Fig. S7 [42]). It corresponds to a YSR state lifetime of $204 \pm 4$ ps which is much shorter than that expected from Fermi-Dirac statistics at our measurement temperature (50 mK). This discrepancy was also reported in the previous studies of YSR lifetimes [44, 43] emphasizing the need for experimental methods that can quantify the transport dynamics. In this regard, SRS offers the possibility that the dynamics of YSR states can be measured without changing the junction conductance or bias voltage, either of which was necessary in previous work [44, 43]. This enables future experiments in which the influence of the tip on the YSR state may be resolved or even leveraged to tune other key parameters such as the YSR state energy.

The results on these three different systems characterize transition rates of individual quantum spins, Néel states, as well as quasiparticle transport dynamics through YSR states. This highlights that SRS measures transition rates with very high accuracy irrespective of the physical origin of the investigated dynamics. As such, it is a valuable addition to the set of measurement techniques that are available for investigating atomic-scale systems. In the next section, we describe the theoretical framework from which eq.1 is derived (section III). Subsequently, we go into detail on the experimental requirements for obtaining high quality SRS spectra (section IV) and then, we show how SRS can be used to gain experimental insight into the energy dependence of transition rates (section V and VI).

## III. ANALYTICAL THEORY OF STOCHASTIC RESONANCE SPECTROSCOPY

The SRS signal in Fig. 1 can be understood from the behavior of a general two-level system. Starting with requirement (i), we modulate the rates of the system using the bias voltage and observe that the two-level system responds differently at low vs. high frequencies. When this external modulation is at a frequency $f$ that is much lower than the transition rates ($f \ll \Gamma/2\pi$) , the dynamics vary smoothly with the drive. In this slow modulation limit (Fig. 2b, left panel), the switching events (gray) happen on a much faster timescale than the modulation and individual switches are not synchronized with the voltage (yellow). Yet, the system's response can be described by a slow oscillation of the steady-state occupation (blue) at the drive frequency. Conversely, when the drive frequency is much faster than the transition rates ($f \gg \Gamma/2\pi$), multiple periods of the external modulation pass between each transition. In this fast-modulation limit (Fig. 2b right panel), the system's response is dominated by random switching according to time-averaged transition rates: the amplitude oscillation of the state occupation is reduced and eventually vanishes for $f \to \infty$. The cross-over between these two regimes is the origin of the characteristic lineshape of SRS (Fig. 2b middle panel), which can be detected using the state-dependent observable (requirement (ii)).

An analytic expression for the SRS signal can be obtained using Markovian transition rates. The harmonic drive voltage modulates the excitation $\Gamma_+$ and relaxation $\Gamma_-$ rates, which are thus time- and frequency-dependent: $\Gamma_\pm(V) = \Gamma_\pm(V_{dc} + V_{ac} \cdot \cos(2\pi f t))$. As a result, the occupation probabilities have the same periodicity as the drive (Fig. 2b).The occupation probability of the excited state can be expressed as a Fourier series $n(t) = \sum n_q(f) e^{2iq\pi f t}$. The tunnel current contains a product of both the bias voltage (yellow) and the occupation probabilities (blue): $I = (V_{dc} + V_{ac} \cdot \cos(2\pi f t))(\sigma_g + \Delta\sigma \sum n_q(f) e^{2\pi i q f t})$, where $\sigma_g$ is the conductance of the ground state, and $\Delta\sigma$ is the difference in conductance between excited and ground state (Supplemental Material, section S1 [42]).

This gives us two main contributions to the SRS signal: the average occupation ($q = 0$ term) weighted by $V_{dc}$, and the first harmonic ($q = \pm 1$ terms) which down-mix with $V_{ac}$ . Both result in a tunnel current that is time-independent and therefore detectable with a conventional amplifier, even if the state occupation itself oscillates at a higher frequency. It is worth noting that the SRS signal is often dominated by the down-mixed component which can be understood as homodyne detection of the occupation [46]. Higher harmonic terms with $|q| > 1$ oscillate at multiples of the drive frequency and usually do not mix to a detectable tunnel current.

Using the two-state model, the occupation probability of both states can be obtained by solving two integrals that depend on the excitation rate and the total transition rate [20]. Using the Fourier series of $\Gamma(t)$ and $\Gamma_+(t)$ it is possible to solve these integrals analytically and obtain the exact time-dependent occupation probability for arbitrary transition rates (Supplemental Material, section S1 [42]). For many physical systems, SRS is well-described when neglecting variations of the total transition rate over time, i.e. $\Gamma(t) = \Gamma$ is constant. In this case, the occupation probability is given by

$$n(t) = \sum_q \frac{a_q}{\Gamma} \frac{1 - iq\Omega}{q^2\Omega^2 + 1} e^{iq\Omega\Gamma t}, \tag{2}$$

where $a_q$ are the Fourier coefficients of the excitation rate and $q = 0, \pm 1, \pm 2, \ldots$ . Eq. 2 directly quantifies the qualitative description discussed above (Fig. 2): at low frequencies, $\Omega \ll 1$, the solution reduces to the adiabatic steady-state solution $n(t) = \Gamma_+(t)/\Gamma$ in which the switching follows the drive oscillation; and at high frequencies, $\Omega \gg 1$, eq. 2 approaches $n(t) = a_0/\Gamma = \langle \Gamma_+(t) \rangle / \Gamma$, which is not time-dependent and describes the loss of synchronization with the drive oscillation[1]. The transition between these low and high frequency limits follows a Fourier series in which the $q^{\text{th}}$ harmonic of the drive frequency has an amplitude $a_q/\Gamma$ that decays as $1/\sqrt{q^2\Omega^2+1}$ and has a frequency-dependent phase shift $\arctan(q\Omega)$. This yields an in-phase component of the occupation probability with a Lorentzian decay $\propto 1/(q^2\Omega^2 + 1)$.

Eq. 2 also specifies why the homodyne component is typically the dominant contribution to the SRS signal. For constant Γ, the time-averaged occupation probability ($q = 0$) is given by $n_0 = a_0/\Gamma$ and does not depend on the frequency. It creates a constant offset without a step in frequency that is included in $I_0$. By contrast, the homodyne signal ($q = \pm 1$) is given by the in-phase part of the first harmonic $n_{\pm 1} = a_{\pm 1}/\Gamma \cdot 1/(\Omega^2 + 1)$ and produces the characteristic step-like lineshape of SRS

$$I^{\Gamma=const.} = \Delta\sigma V_{ac} \frac{a_{\pm 1}}{\Gamma} \frac{1}{\Omega^2 + 1} + I_0 \,. \tag{3}$$

We find that eq. 3 gives the lineshape of eq. 1 and quantifies the SRS magnitude in terms of the Fourier components of the transition rates. The width of the step is universal and independent of the details of the investigated system. The height of the step scales with the drive amplitude $V_{ac}$, the slope of the Γ(V) dependence ($\propto a_{\pm 1}$), the difference in conductivity of the states $\Delta\sigma$ and is inversely proportional to the transition rate Γ. As a result, the SRS signal amplitude, $I_{SR} = \Delta\sigma V_{ac} a_{\pm 1}/\Gamma$, can have either sign, depending on the bias polarity and the sign of $\Delta\sigma$ and $a_{\pm 1}$. Additional information about the physics of the investigated system can be extracted from analyzing its energy dependence (see sections V and VI).

It is worth noting that these results, which are derived for a constant total rate, $\Gamma(t) = const.$, remain valid as long as the system stays in a low modulation regime where the amplitude of the drive is low enough to keep the modulation of the rates small compared to the intrinsic rates. This is the case for a large range of systems, as exemplified in Fig. 1.

Beyond the low modulation regime, higher order contributions to the SRS signal can become relevant as they may distort the Lorentzian lineshape of the SRS signal, as displayed in the Supplemental Material Fig. S7c-f [42]. The higher order terms are caused by a mixing of different harmonics which gives rise to a series of terms where the $j^{th}$ term scales with $O\left(\frac{a_k}{\Gamma}\prod_{i=1}^{j-1}\frac{b_{m_i}}{\Gamma}\right)$, where $k = q - \sum_{i=1}^{j-1} m_i$, and $\{m_i\}$ is a set of $j-1$ integers (details in Supplemental Material, section S1 [42]). These terms have the same Lorentzian shape as the terms in eq. 2. but may be shifted towards lower frequencies. This can occur for example, under specific drive conditions where the excitation and relaxation rates modulate strongly with the drive such that $\Gamma(t)$ deviates significantly from being constant. Measurements in this higher order regime are identifiable by visible deviations of the measured signal from the normal SRS lineshape as defined in eq. 1 (Supplemental Material Fig. S7a,b [42]).

---

[1] Both limits can also be obtained for time-dependent Γ(t) (Supplemental Material, section S2 [42]).

Interestingly, the first Fourier components ($|k| \leq 1$) of the 2nd order terms are at the same frequency as the 1st order terms, i.e. the step-like lineshape of SRS remains preserved and only its amplitude is affected. As a consequence, the SRS lineshape stays preserved up to moderately large modulations of the total rate. We find that linear modulations of $\Gamma(t)$ up to 60% lead to an underestimation of $\Gamma$ of only 5% (See Supplemental Material Fig. S8 [42]). This is comparable to the experimental noise of a typical SRS measurement and therefore an acceptable deviation. More stringent limitations apply to systems with a highly non-linear drive dependence, e.g. when the modulation crosses an energy threshold. In the section V we address behavior at such energy thresholds.

## IV. EXPERIMENTAL IMPLEMENTATION OF STOCHASTIC RESONANCE SPECTROSCOPY

Fundamentally, the experimental implementation of SRS requires driving by a periodic modulation of one experimental input that affects the transition rates of the investigated system and recording of one experimental output that depends on the system's state occupation. For STM, the experimental output is typically the tunnel current and the most convenient experimental input is the bias voltage because many atomic-scale systems can be excited by tunneling electrons when the bias provides them with sufficient energy. Additionally, the bias voltage in the tunnel junction can be modulated at near-arbitrary frequencies ranging from DC to several THz [29, 47, 48]. In this work, we use electronic frequency sources that operate up to the microwave range.

To separate the stochastic resonance spectroscopy signal (SRS signal) produced by the bias voltage modulation from other components of the tunnel current that are drive-independent, we chop the microwave modulation on and off at 691 Hz and detect the response of the tunnel current using a lock-in amplifier. The resulting lock-in signal contains the SRS signal and the rectified tunnel current. The rectified current originates from rectification of the microwave by non-linear $I(V)$ characteristics, which simply adds a frequency-independent offset. The measured SRS signal follows the universal Lorentzian lineshape of eq. 1 and 3. Since the width of the SRS signal is universal, the sampling of the SRS signal as a function of frequency does not need to be adjusted to the system of interest: the most efficient strategy is to sweep the frequency logarithmically with approximately 30 points per decade. This is sufficient to accurately detect the SRS step position on top of the frequency-independent contributions.

Importantly, the signal-to-noise of SRS can be much higher than real-time detection of stochastic switching or pump probe measurements allow. This is because it is performed purely in the frequency domain and detection of the SRS signal is independent of the speed of the probed dynamics. The main requirement for high signal-to-noise is a high quality characterization of the amplitude transfer function of the STM (Supplemental Material Fig. S5 [42]) [49, 41, 50]. Moreover, SRS is not limited by temperature: the harmonic drive leads to time-dependent occupation of the states also for arbitrary initial occupation probabilities. Thus, an SRS signal will be detectable even at elevated temperatures, where thermal broadening impedes energy resolution and relaxometry measurements.

The instrumentation requirements of SRS are similar to those of continuous wave electron spin resonance (ESR) spectroscopy in the STM. Both techniques add a sine-wave modulation to the bias voltage and measure the tunneling current while sweeping the drive frequency. But the underlying processes are fundamentally different: ESR-STM probes coherent dynamics between states and the frequency at which the ESR signal occurs translates to the excitation energy between the states [51]. Sweeping the drive frequency in a narrow window allows determining the excitation energy with very high resolution [51]. By contrast, SRS exploits incoherent dynamics and measures the transition rates between states. The origin of the SRS signal lies in the frequency dependence of the state occupation. Consequently, the SRS signal spans a much larger frequency range that typically covers several orders of magnitude. As such, SRS measurements are analogous to AC susceptibility measurements [52, 53, 54], real-time measurements of telegraph noise [6, 7, 8, 9, 10, 11, 13, 12] or relaxometry methods such as pump-probe spectroscopy [27, 28, 29, 30].

The broad applicability and system-agnostic nature of SRS makes it an efficient screening tool for identifying suitable candidates for ESR-STM (where linewidth limits resolution) and for optimizing pump-probe experimental parameters across different lifetime regimes. Additionally, SRS provides complementary information to pump-probe spectroscopy by measuring transition rates under continuous driving, enabling characterization of driven quantum dynamics rather than just relaxation processes (see section VI for details on driven dynamics).

## V. ENERGY DEPENDENCE OF TRANSITION RATES

SRS probes the transition rate of a system and therefore can help determining how its dynamics are affected by interactions with the environment. In an STM, the most direct way to tune this interaction is by applying a DC bias voltage across the tunnel junction. This is particularly relevant because the energy of the tunneling electrons determines which transitions are possible in the studied system. However, so far these interactions with the environment and the resulting changes in transition rates have only been accessible through simulation [55, 56].

We consider a general scenario with a distinct change of the excitation rate $\Gamma_+(V)$ at a threshold bias voltage $V_{th}$ , Fig. 3a (yellow). Below the threshold, the excitation rate increases only weakly with applied voltage, while above the threshold, it increases strongly with excess voltage $|V - V_{th}|$. Such behavior occurs, for example, due to inelastic electron tunneling where tunneling electrons can excite transitions to higher excited states once a threshold voltage is surpassed, which opens additional transition pathways. To understand the SRS signal at such energy thresholds, it is sufficient to consider the case of constant relaxation rate[2] (Fig. 3a, blue). Here, we consider the low modulation regime discussed in the previous section by setting the excitation rate to $\Gamma_+ < 0.3\,\Gamma$. We simulate SRS measurements numerically following a rate equation approach as described in Supplemental Material, section S3 [42]. The simulated SRS signal is then fitted with eq. 1 to extract the total transition rate and the SRS signal amplitude and the obtained values are compared to the analytical solution of the same system.

We first investigate the impact of the modulation amplitude $V_{ac}$ on the signal amplitude (Fig. 3b). The fitted numerical SRS amplitude (black dots) increases approximately quadratically once $V_{ac} + V_{dc}$ crosses the threshold voltage $V_{th}$ (red line). This behavior can be understood using eq. 3 where the amplitude of the SRS signal is given by $I_{SR} = \Delta\sigma\, V_{ac} \cdot a_{\pm 1}/\Gamma$. Below the threshold, the excitation rate is very weakly modulated, i.e. $a_{\pm 1}$ is small; this results in a small amplitude of the SRS signal. Once the voltage reaches the threshold ($V_{dc} + V_{ac} > V_{th}$), this modulation is much stronger and increases linearly with excess voltage. The SRS amplitude, being proportional to the product of $a_{\pm 1}$ and $V_{ac}$, scales with $V_{ac}(V_{ac} + V_{dc} - V_{th})$, which gives rise to the observed quadratic behavior. Increasing the drive amplitude is thus a good strategy to improve the signal-to-noise ratio of the measurements.

[2] See Supplemental Material Fig. S9 [42] for results with $\Gamma_-(t) \neq const.$.

Yet, we note that large modulation amplitudes lead to quantitative deviation from eq. 3 once the low modulation regime is left. Thus, higher order terms need to be taken into account to recover a quantitative agreement. For the system studied here 2$^{nd}$ order terms must be included in the analytical solution for modulation amplitudes above 2 mV (Fig. 3b dark blue curve) to reproduce the full numerical results, because the 1$^{st}$ order analytical solution (light blue curve) overestimates the SRS signal. The qualitative behavior is nonetheless well-captured by eq. 3. Importantly, as shown in Supplemental Material Fig. S10c,d [42], the main contribution to these deviations are the terms with $j = 2$ and $|k| \leq 1$ which do not distort the SRS signal, i.e. the transition rate in this regime is still measured accurately.

This robustness to large modulation makes SRS a powerful technique to probe the energy dependence of the system's dynamics across energy thresholds. We investigate the dependence of the total rate with offset voltage $V_{dc}$ while keeping $V_{ac}$ constant (Fig. 3c). Upon reaching the threshold (red line), the total rate transitions from a slow increase in the sub-threshold regime to a steeper increase above the threshold. Above and below the threshold, the rate obtained from fitting the numerical simulation (dots) agrees with the real input rate to within 0.5%. Modulating the voltage over the threshold (gray areas) slightly increases the error, but it stays below 2%, which means the relative error would remain below typical measurement uncertainties in an experiment. This highlights that varying SRS can quantitatively measure the energy dependence of the transition rate as function of bias and even across energy thresholds.

The SRS signal amplitude also contains information about the transition rates that helps characterize the crossing of a threshold. As depicted in Fig. 3d, such a crossing results in a linear increase of the signal amplitude followed by a slow decay. This behavior originates in the evolution of the rate modulation and the total rate across the threshold: as the voltage crosses the threshold, the larger slope of the excitation rate increases the modulation of the rates i.e. $a_{\pm 1}$ in eq. 3. The signal amplitude thus rises until the bias voltage remains fully above the threshold, $|V_{dc}| + |V_{ac}| > |V_{th}|$. A similar behavior is seen at $V_{dc} = 0$ due to the inversion of the slope of $\Gamma_{+}$. Beyond the threshold, the modulation of the rates and thus $a_{\pm 1}$ stays constant while the total rate $\Gamma$ keeps increasing. As a result, the signal decreases with $1/\Gamma$. This trend shown by the numerical simulation (Fig. 3d, black dots) is qualitatively captured in the analytical model, even to first order (light blue curve). For a quantitative agreement, it is necessary to consider higher order processes past the energy threshold (2$^{nd}$ order shown in dark blue) analogous to what was observed in Fig. 3b.

In total, an increase in the slope of the excitation rate causes a sharp increase in the signal amplitude, while an increase in the total rate reduces the signal amplitude. In addition, the SRS signal typically reverses sign with bias polarity, such that negative SRS signal at positive bias becomes a positive signal at negative bias and vice versa. This characteristic behavior can be used to detect thresholds and thereby identify different transition pathways in multi-state systems as exemplified in the following section.

## VI. CHARACTERIZING DRIVEN AND FREE DYNAMICS OF MULTISTATE SYSTEM

The previous section highlights that the transition rates of a system are highly dependent on how it is driven. Understanding and accessing these driven dynamics is crucial, for example, in quantum computing and quantum sensing where qubits must remain stable while gate operations are applied. Hence, both the intrinsic and extrinsic relaxation rates have to be characterized and optimized to enhance the qubits' lifetimes. Here we show that SRS measurements allow measuring and quantifying both the driven and undriven dynamics of a single Fe atom on a Cu binding site on $Cu_2N$.

The Fe atom is an effective spin $S = 2$ system with easy axis anisotropy and with an known energy diagram, as previously characterized [40]. In a magnetic field its two lowest-energy spin states with $m_s = \pm 2$ are separated by an energy barrier comprising the $m_s = \pm 1$ and $m_s = 0$ states. As a result, transitions can occur either as multi-step processes through higher excited states or as a direct transition through tunneling of magnetization. Which of these pathways is dominant depends on the offset voltage, $V_{dc}$, and we show that SRS can resolve the crossover between them and track the changes in the Fe spin's dynamics as they are tuned with the bias voltage.

Figure 4a shows exemplary SRS measurements (dots) recorded at offset voltages of $V_{dc} = +2\,mV$, $-2\,mV, -6\,mV$. All three measurements display the characteristic step that can be fitted with the analytical SRS signal of eq. 1 (solid lines). For $V_{dc} = \pm 2\,mV$, the steps are at a similar frequency, yielding a transition rate of $\Gamma(\pm 2mV) = (1.4 \pm 0.1) \cdot 10^9/s$ , but the SRS signal has opposite sign. At $V_{dc} = -6\,mV$, the SRS signal has the same sign as at $-2\,mV$ but with a slightly larger amplitude. Importantly, the step is shifted to higher frequency, indicating a larger total transition rate ($\Gamma(-6mV) = (1.78 \pm 0.05) \cdot 10^9/s$ ). As discussed in the previous section, this increase in signal amplitude is an indication of a larger slope in the excitation rate, pointing toward the presence of an energy threshold.

To gain further insights into the energy dependence of the Fe atom's spin dynamics, we evaluate the variation of the total rate and the signal amplitude as a function of offset bias voltage (Fig. 4b,c). For small offset voltages ($eV_{dc} < E_1$), where $E_1$ is energy of the 3[rd] excited state, we observe a small SRS signal that increases in magnitude with bias offset. At the same time, $\Gamma$ remains constant within the measurement accuracy. This indicates that the modulation of the transition rates by the external drive is negligible compared to the intrinsic, bias-independent transition rate of the Fe spin caused by tunneling of magnetization. The measured transition rate is thus that of the unperturbed or undriven system.

This behavior differs markedly for higher offset voltages: we observe a linear increase of $\Gamma$ with two different slopes for $E_1 < e|V_{dc} \pm V_{ac}| < E_2$ and $e|V_{dc} \pm V_{ac}| > E_2$, where $E_1 \approx 4\,meV$ and $E_2 \approx 6\,meV$ correspond to two possible spin transitions in the Fe atom [40]. In parallel, the SRS signal amplitude increases at $E_1$ and decreases at $E_2$. The opposite signs of the SRS amplitude at positive and negative bias polarities are due to a reversal of the slope of the excitation rate, as explained in the previous section. The difference in signal amplitude between the two polarities indicates different absolute slopes of the excitation rate at positive and negative bias voltage, which relates to the spin polarization of the transitions also leading to asymmetric spin-polarized dI/dV spectra (Supplemental Material Fig. S2 [42]). Around $V_{dc} = +5\,mV$ the signal amplitude vanishes and the experimental data cannot be fitted.

The total transition rate of the system results from the addition of all allowed pathways. This is indicated by the shaded areas in Fig. 4b showing their relative weights: for instance, at $-5.5\ mV$, driven dynanics via the first pathway contribute to 15% of the total dynamics, and at $-9\ mV$, the driven dynamics contain contributions from both the first (30%) and the second (15%) pathway. Moreover, the increase of the signal amplitude at $E_1$ further indicates a strong increase in the slope of the excitation rate of the system while the decrease at $E_2$ is related to the increase of the total relaxation rate and indicates a weak or vanishing increase in the slope of the excitation rate.

With the additional knowledge of the spin transition energies by inelastic electron tunneling spectroscopy (Supplemental Material Fig. S2 [42]) [57], we can form a complete picture of how the transition pathways in the Fe atom change and which quantum states are involved in the dynamics as a function of offset voltage. The inset in Fig. 4c shows that the first threshold $E_1$ corresponds to a transition between the $m_s = -2$ ground state and $m_s = -1$ excited state, and the second threshold $E_2$ to that between $m_s = 2$ and $m_s = 1$. As a result, we can identify the transition pathways that cause the two thresholds. At the first threshold, excitations into the $m_s = -1$ excited state can decay into the $m_s = 2$ state (red). This path acts as an excitation and thus increases the SRS signal amplitude. At the second threshold, the transitions into the $m_s = 1$ state can decay into the ground state (blue). This constitutes a relaxation and predominantly increases the total rate, while decreasing the signal amplitude.

Indeed, a numerical simulation of SRS measurements analogous to that used in the previous section but using established scattering theory for magnetic atoms on surfaces [55, 56] and a Master rate equation for all five spin states matches well to the experimental results and verifies the voltage-dependent crossover between the different transition pathways (Supplemental Material Fig. S12 [42]). Note that the scattering rates with the sample bath are very high for this system.

Having identified the different transition pathways, we can quantify their efficiency and contribution to the driven dynamics by comparing the slope of the tunnel current, $I(V)$, with that of $\Gamma(V)$. From the $I(V)$ characteristics we can determine the rate of tunneling electrons that could excite the Fe spin: between $E_1/e$ and $E_2/e$ the tunneling rate increases by $\sim 1.3 \cdot 10^9\ s^{-1} mV^{-1}$. In the same voltage interval $\Gamma$ increases only by $0.15 \cdot 10^9\ s^{-1} mV^{-1}$ (Fig. 4b) which indicates that 10% of the tunneling electrons participate in the driven dynamics. Likewise, for a bias surpassing $E_2/e$, the driven dynamics – that now stem from both pathways – have a 35% efficiency ($I(V)$ increases by $0.8 \cdot 10^9\ s^{-1} mV^{-1}$, whereas Γ(V) increases by $0.28 \cdot 10^9 s^{-1} mV^{-1}$).

Summarizing this analysis, we find that characterizing the bias dependence of the SRS signal allows accessing both driven and free dynamics of the Fe atom's spin and makes the crossover of different transition pathways between the spin states observable. We believe, that this methodology is applicable to other multi-level systems and not limited to spin.

## VII. CONCLUSION

In summary, we have shown that SRS enables quantitative measurements of both spin and transport dynamics at the atomic scale. It resolves transition rates over a wide frequency range, spanning from kilohertz to gigahertz. This spectroscopy method only requires two conditions to be fulfilled: (i) the transition probabilities between the investigated states of the system must depend on an external control parameter that can be modulated at different frequencies and (ii) it must be possible to record the time average of an observable that depends on the state of the investigated system.

We present a theoretical framework that allows for quantitative determination of the transition rates from SRS measurements irrespective of the details of the studied system. In particular, we show that SRS can be used to study the energy dependence of the transition rates, which were so far only accessible through simulations. This sheds light onto the interaction of a discrete system with its environment. Finally, SRS can be used in multi-level systems to detect the onset of different transition pathways. This enables the investigation of both the free and driven dynamics in such systems. An exciting next step will be the investigation of systems that display more complex behavior, arising from coherent dynamics (either intrinsic or induced by a drive) or from correlations with the environment leading to non-Markovian behavior.

We expect that SRS can be used to investigate the dynamics of a wide variety of other discrete systems such as the dynamics of orbitals, charge states, quasi-particle lifetimes in atoms, molecules or other low-dimensional defects, and ultrafast dynamics emerging from strong correlations with a bath [58]. Shifting away from observing dynamics via the tunneling current, SRS may also be implemented using luminescence as the observable, enabling the measurement of localized exciton dynamics in an STM. The universality of stochastic processes further makes SRS potentially applicable to other research fields that go beyond STM investigations, for example studies in quantum dots [18]. Additionally, SRS bears similarity to AC susceptibility measurements that are frequently used in the characterization of ensembles of molecular magnets and molecular qubits [52, 53, 54]. We therefore anticipate that SRS will also allow for new types of single-molecule studies on surfaces and open up possibilities for the investigation of coupled spins or molecular systems.

## ACKNOWLEDGEMENTS

N.B. acknowledges support from the Carl-Zeiss-Stiftung, the Centre for Integrated Quantum Science and Technology ($IQ^{ST}$). L.F. acknowledges support from the Carl-Zeiss-Stiftung, CZS Center QPhoton. S.N.C. acknowledges support from the US Army Research Office, Award #W911NF-23-1-0110 and from the Australian Research Council, Project No. DP210101608. S.B. and S.L. acknowledge support from the Baden Württemberg Foundation Program on Quantum Technologies (Project AModiQuS).

## REFERENCES

[1] M. B. Plenio and P. L. Knight, "The quantum-jump approach to dissipative dynamics in quantum optics," *Rev. Mod. Phys.,* vol. 70, pp. 101-144, January 1998.

[2] U. Seifert, "Stochastic thermodynamics, fluctuation theorems and molecular machines," *Rep. Prog. Phys.,* vol. 75, p. 126001, November 2012.

[3] G. Giusi, G. Cannatà, G. Scandurra and C. Ciofi, "Ultra-low-noise large-bandwidth transimpedance amplifier," *Int. J. Circ. Theor. Appl.,* vol. 43, p. 1455–1473, August 2014.

[4] A. V. der Ziel, Noise in solid state devices and circuits, New York: Wiley, 1986.

[5] R. Gao, M. A. Edwards, J. M. Harris and H. S. White, "Shot noise sets the limit of quantification in electrochemical measurements," *Curr. Opin. Electrochem.,* vol. 22, p. 170–177, August 2020.

[6] S. Krause, L. Berbil-Bautista, G. Herzog, M. Bode and R. Wiesendanger, "Current-Induced Magnetization Switching with a Spin-Polarized Scanning Tunneling Microscope," *Science,* vol. 317, p. 1537–1540, September 2007.

[7] L. Gerhard, T. K. Yamada, T. Balashov, A. F. Takács, R. J. H. Wesselink, M. Däne, M. Fechner, S. Ostanin, A. Ernst, I. Mertig and W. Wulfhekel, "Magnetoelectric coupling at metal surfaces," *Nat. Nanotechnol.,* vol. 5, p. 792–797, October 2010.

[8] S. Loth, S. Baumann, C. P. Lutz, D. M. Eigler and A. J. Heinrich, "Bistability in Atomic-Scale Antiferromagnets," *Science,* vol. 335, p. 196–199, January 2012.

[9] A. A. Khajetoorians, B. Baxevanis, C. Hübner, T. Schlenk, S. Krause, T. O. Wehling, S. Lounis, A. Lichtenstein, D. Pfannkuche, J. Wiebe and R. Wiesendanger, "Current-Driven Spin Dynamics of Artificially Constructed Quantum Magnets," *Science,* vol. 339, p. 55–59, January 2013.

[10] A. Spinelli, B. Bryant, F. Delgado, J. Fernández-Rossier and A. F. Otte, "Imaging of spin waves in atomically designed nanomagnets," *Nat. Mater.,* vol. 13, p. 782–785, July 2014.

[11] K. Teichmann, M. Wenderoth, S. Loth, J. K. Garleff, A. P. Wijnheijmer, P. M. Koenraad and R. G. Ulbrich, "Bistable Charge Configuration of Donor Systems near the GaAs(110) Surfaces," *Nano Lett.,* vol. 11, p. 3538–3542, August 2011.

[12] W. Paul, K. Yang, S. Baumann, N. Romming, T. Choi, C. Lutz and A. Heinrich, "Control of the millisecond spin lifetime of an electrically probed atom," *Nat. Phys.,* vol. 13, p. 403–407, November 2016.

[13] B. Kiraly, A. N. Rudenko, W. M. J. van Weerdenburg, D. Wegner, M. I. Katsnelson and A. A. Khajetoorians, "An orbitally derived single-atom magnetic memory," *Nat. Commun.,* vol. 9, p. 3904, September 2018.

[14] S. Fatayer, F. Albrecht, I. Tavernelli, M. Persson, N. Moll and L. Gross, "Probing Molecular Excited States by Atomic Force Microscopy," *Phys. Rev. Lett.,* vol. 126, p. 176801, April 2021.

[15] M. Hänze, G. McMurtrie, S. Baumann, L. Malavolti, S. N. Coppersmith and S. Loth, "Quantum stochastic resonance of individual Fe atoms," *Sci. Adv.,* vol. 7, no. eabg2616, August 2021.

[16] W. M. J. van Weerdenburg, H. Osterhage, R. Christianen, K. Junghans, E. Domínguez, H. J. Kappen and A. A. Khajetoorians, "Stochastic Syncing in Sinusoidally Driven Atomic Orbital Memory," *ACS Nano,* vol. 18, p. 4840–4846, January 2024.

[17] J. Yao, S. Chen, W. Shi and W. Ho, "Quantum Stochastic Rectification in a Single Molecule," *Physical Review Letters,* vol. 135, July 2025.

[18] T. Wagner, P. Talkner, J. C. Bayer, E. P. Rugeramigabo, P. Hänggi and R. J. Haug, "Quantum stochastic resonance in an a.c.-driven single-electron quantum dot," *Nat. Phys.,* vol. 15, p. 330–334, February 2019.

[19] B. McNamara and K. Wiesenfeld, "Theory of stochastic resonance," *Phys. Rev. A,* vol. 39, p. 4854–4869, May 1989.

[20] R. Löfstedt and S. N. Coppersmith, "Stochastic resonance: Nonperturbative calculation of power spectra and residence-time distributions," *Phys. Rev. E,* vol. 49, p. 4821–4831, June 1994.

[21] L. Gammaitoni, P. Hänggi, P. Jung and F. Marchesoni, "Stochastic resonance," *Rev. Mod. Phys.,* vol. 70, p. 223–287, January 1998.

[22] R. Benzi, G. Parisi, A. Sutera and A. Vulpani, "Stochastic resonance in climatic change," *Tellus,* vol. 34, p. 10–16, February 1982.

[23] S. Fauve and F. Heslot, "Stochastic resonance in a bistable system," *Physics Letters A,* vol. 97, p. 5–7, August 1983.

[24] A. Longtin, A. Bulsara and F. Moss, "Time-interval sequences in bistable systems and the noise-induced transmission of information by sensory neurons," *Physical Review Letters,* vol. 67, p. 656–659, July 1991.

[25] S. Loth, K. von Bergmann, M. Ternes, A. F. Otte, C. P. Lutz and A. J. Heinrich, "Controlling the state of quantum spins with electric currents," *Nat. Phys.,* vol. 6, p. 340–344, March 2010.

[26] J. Hermenau, M. Ternes, M. Steinbrecher, R. Wiesendanger and J. Wiebe, "Long Spin-Relaxation Times in a Transition-Metal Atom in Direct Contact to a Metal Substrate," *Nano Lett.,* vol. 18, p. 1978–1983, February 2018.

[27] S. Loth, M. Etzkorn, C. P. Lutz, D. M. Eigler and A. J. Heinrich, "Measurement of Fast Electron Spin Relaxation Times with Atomic Resolution," *Science,* vol. 329, pp. 1628-1630, 2010.

[28] C. Saunus, J. Raphael Bindel, M. Pratzer and M. Morgenstern, "Versatile scanning tunneling microscopy with 120 ps time resolution," *Appl. Phys. Lett.,* vol. 102, p. 051601, February 2013.

[29] T. L. Cocker, V. Jelic, M. Gupta, S. J. Molesky, J. A. J. Burgess, G. D. L. Reyes, L. V. Titova, Y. Y. Tsui, M. R. Freeman and F. A. Hegmann, "An ultrafast terahertz scanning tunnelling microscope," *Nat. Photonics,* vol. 7, p. 620–625, July 2013.

[30] S. Yoshida, Y. Aizawa, Z.-h. Wang, R. Oshima, Y. Mera, E. Matsuyama, H. Oigawa, O. Takeuchi and H. Shigekawa, "Probing ultrafast spin dynamics with optical pump–probe scanning tunnelling microscopy," *Nat. Nanotechnol.,* vol. 9, p. 588–593, June 2014.

[31] K. Yang, W. Paul, S.-H. Phark, P. Willke, Y. Bae, T. Choi, T. Esat, A. Ardavan, A. J. Heinrich and C. P. Lutz, "Coherent spin manipulation of individual atoms on a surface," *Science,* vol. 366, p. 509–512, October 2019.

[32] Y. Wang, Y. Chen, H. T. Bui, C. Wolf, M. Haze, C. Mier, J. Kim, D.-J. Choi, C. P. Lutz, Y. Bae, S.-h. Phark and A. J. Heinrich, "An atomic-scale multi-qubit platform," *Science,* vol. 382, p. 87–92, October 2023.

[33] J. Doležal, A. Sagwal, R. C. de Campos Ferreira and M. Švec, "Single-Molecule Time-Resolved Spectroscopy in a Tunable STM Nanocavity," *Nano Lett.,* vol. 24, p. 1629–1634, January 2024.

[34] P. Merino, C. Große, A. Rosławska, K. Kuhnke and K. Kern, "Exciton dynamics of C60-based single-photon emitters explored by Hanbury Brown–Twiss scanning tunnelling microscopy," *Nat. Commun.,* vol. 6, September 2015.

[35] K. Jhuria, J. Hohlfeld, A. Pattabi, E. Martin, A. Y. Arriola Córdova, X. Shi, R. Lo Conte, S. Petit-Watelot, J. C. Rojas-Sanchez, G. Malinowski, S. Mangin, A. Lemaître, M. Hehn, J. Bokor, R. B. Wilson and J. Gorchon, "Spin–orbit torque switching of a ferromagnet with picosecond electrical pulses," *Nat. Electron.,* vol. 3, p. 680–686, October 2020.

[36] Y. Chew, T. Tomita, T. P. Mahesh, S. Sugawa, S. de Léséleuc and K. Ohmori, "Ultrafast energy exchange between two single Rydberg atoms on a nanosecond timescale," *Nat. Photonics,* vol. 16, p. 724–729, August 2022.

[37] L. U. H. Yu, "Bound state in superconductors with paramagnetic impurities," *Acta Physica Sinica,* vol. 21, p. 75, 1965.

[38] H. Shiba, "Classical Spins in Superconductors," *Prog. Theor. Phys.,* vol. 40, p. 435–451, September 1968.

[39] A. I. Rusinov, "Theory of gapless superconductivity in alloys containing paramagnetic impurities," *Sov. Phys. JETP,* vol. 29, p. 1101–1106, 1969.

[40] C. F. Hirjibehedin, C.-Y. Lin, A. F. Otte, M. Ternes, C. P. Lutz, B. A. Jones and A. J. Heinrich, "Large Magnetic Anisotropy of a Single Atomic Spin Embedded in a Surface Molecular Network," *Science,* vol. 317, p. 1199–1203, August 2007.

[41] S. Baumann, G. McMurtrie, M. Hänze, N. Betz, L. Arnhold, L. Malavolti and S. Loth, "An Atomic-scale Vector Network Analyzer," *Small Methods,* vol. 8, no. 2301526, 2024.

[42] *See Supplemental Material at [] for the full derivation of the SRS signal, the time-dependent occupation in the case of arbitrary transition rates, a table detailing the simulation parameters used in the numerical model, a pump-probe spectrum of a single Fe atom, a differential conductance measurement of a single Fe atom using a spin-polarized tip, plots detailing deviations in the SRS signal from the lineshape of eq. 1, the error in the fitted rate, caused by deviations in the lineshape, details on the contribution of higher order terms to the amplitudes, displayed in Fig. 3 and an estimate of the YSR lifetime as conducted in* [43].

[43] M. Ruby, F. Pientka, Y. Peng, F. von Oppen, B. W. Heinrich and K. J. Franke, "Tunneling Processes into Localized Subgap States in Superconductors," *Phys. Rev. Lett.,* vol. 115, p. 087001, August 2015.

[44] H. Huang, C. Padurariu, J. Senkpiel, R. Drost, A. L. Yeyati, J. C. Cuevas, B. Kubala, J. Ankerhold, K. Kern and C. R. Ast, "Tunnelling dynamics between superconducting bound states at the atomic limit," *Nat. Phys.,* vol. 16, p. 1227–1231, July 2020.

[45] U. Thupakula, V. Perrin, A. Palacio-Morales, L. Cario, M. Aprili, P. Simon and F. Massee, "Coherent and Incoherent Tunneling into Yu-Shiba-Rusinov States Revealed by Atomic Scale Shot-Noise Spectroscopy," *Phys. Rev. Lett.,* vol. 128, p. 247001, June 2022.

[46] Y. Bae, K. Yang, P. Willke, T. Choi, A. J. Heinrich and C. P. Lutz, "Enhanced quantum coherence in exchange coupled spins via singlet-triplet transitions," *Sci. Adv.,* vol. 4, no. eaau4159, November 2018.

[47] M. Garg and K. Kern, "Attosecond coherent manipulation of electrons in tunneling microscopy," *Science,* vol. 367, p. 411–415, January 2020.

[48] S. Sheng, M. Abdo, S. Rolf-Pissarczyk, K. Lichtenberg, S. Baumann, J. A. J. Burgess, L. Malavolti and S. Loth, "Terahertz spectroscopy of collective charge density wave dynamics at the atomic scale," *Nature Physics,* vol. 20, p. 1603–1608, July 2024.

[49] W. Paul, S. Baumann, C. P. Lutz and A. J. Heinrich, "Generation of constant-amplitude radio-frequency sweeps at a tunnel junction for spin resonance STM," *Rev. Sci. Instrum.,* vol. 87, no. 074703, July 2016.

[50] M. Hervé, M. Peter and W. Wulfhekel, "High frequency transmission to a junction of a scanning tunneling microscope," *Appl. Phys. Lett.,* vol. 107, no. 093101, August 2015.

[51] S. Baumann, W. Paul, T. Choi, C. P. Lutz, A. Ardavan and A. J. Heinrich, "Electron paramagnetic resonance of individual atoms on a surface," *Science,* vol. 350, p. 417, October 2015.

[52] D. Gatteschi, R. Sessoli and J. Villain, Molecular Nanomagnets, Oxford: Oxford University Press, 2006.

[53] M. del Carmen Giménez-López, F. Moro, A. La Torre, C. J. Gómez-García, P. D. Brown, J. van Slageren and A. N. Khlobystov, "Encapsulation of single-molecule magnets in carbon nanotubes," *Nat. Commun.,* vol. 2, p. 407, July 2011.

[54] L. Tesi, A. Lunghi, M. Atzori, E. Lucaccini, L. Sorace, F. Totti and R. Sessoli, "Giant spin–phonon bottleneck effects in evaporable vanadyl-based molecules with long spin coherence," *Dalton Trans.,* vol. 45, p. 16635–16643, 2016.

[55] S. Loth, C. P. Lutz and A. J. Heinrich, "Spin-polarized spin excitation spectroscopy," *New J. Phys.,* vol. 12, p. 125021, December 2010.

[56] M. Ternes, "Spin excitations and correlations in scanning tunneling spectroscopy," *New J. Phys.,* vol. 17, p. 063016, June 2015.

[57] A. J. Heinrich, J. A. Gupta, C. P. Lutz and D. M. Eigler, "Single-Atom Spin-Flip Spectroscopy," *Science,* vol. 306, p. 466–469, October 2004.

[58] J. Kondo, "Resistance Minimum in Dilute Magnetic Alloys," *Prog. Theor. Phys.,* vol. 32, p. 37–49, July 1964.

[59] D. M. Eigler and E. K. Schweizer, "Positioning single atoms with a scanning tunnelling microscope," *Nature,* vol. 344, p. 524–526, April 1990.

[60] F. D. Natterer, K. Yang, W. Paul, P. Willke, T. Choi, T. Greber, A. J. Heinrich and C. P. Lutz, "Reading and writing single-atom magnets," *Nature,* vol. 543, p. 226–228, March 2017.

[61] I. G. Rau, S. Baumann, S. Rusponi, F. Donati, S. Stepanow, L. Gragnaniello, J. Dreiser, C. Piamonteze, F. Nolting, S. Gangopadhyay, O. R. Albertini, R. M. Macfarlane, C. P. Lutz, B. A. Jones, P. Gambardella, A. J. Heinrich and H. Brune, "Reaching the magnetic anisotropy limit of a 3 d metal atom," *Science,* vol. 344, p. 988–992, May 2014.

[62] P. Willke, T. Bilgeri, X. Zhang, Y. Wang, C. Wolf, H. Aubin, A. Heinrich and T. Choi, "Coherent Spin Control of Single Molecules on a Surface," *ACS Nano,* vol. 15, p. 17959–17965, November 2021.

[63] P. Liljeroth, J. Repp and G. Meyer, "Current-Induced Hydrogen Tautomerization and Conductance Switching of Naphthalocyanine Molecules," *Science,* vol. 317, p. 1203–1206, August 2007.

[64] C. Lotze, M. Corso, K. J. Franke, F. von Oppen and J. I. Pascual, "Driving a Macroscopic Oscillator with the Stochastic Motion of a Hydrogen Molecule," *Science,* vol. 338, p. 779–782, November 2012.

[65] D. Binder, E. C. Smith and A. B. Holman, "Satellite Anomalies from Galactic Cosmic Rays," *IEEE Trans. Nucl. Sci.,* vol. 22, p. 2675, 1975.

[66] L. Faoro, A. Kitaev and L. B. Ioffe, "Quasiparticle Poisoning and Josephson Current Fluctuations Induced by Kondo Impurities," *Phys. Rev. Lett.,* vol. 101, p. 247002, December 2008.

[67] L. Gao, Q. Liu, Y. Y. Zhang, N. Jiang, H. G. Zhang, Z. H. Cheng, W. F. Qiu, S. X. Du, Y. Q. Liu, W. A. Hofer and H.-J. Gao, "Constructing an Array of Anchored Single-Molecule Rotors on Gold Surfaces," *Phys. Rev. Lett.,* vol. 101, p. 197209, November 2008.

[68] A. Gover, A. Nause, E. Dyunin and M. Fedurin, "Beating the shot-noise limit," *Nat. Phys.,* vol. 8, p. 877–880, October 2012.

[69] B. W. Heinrich, L. Braun, J. I. Pascual and K. J. Franke, "Protection of excited spin states by a superconducting energy gap," *Nat. Phys.,* vol. 9, p. 765–768, November 2013.

[70] B. Kiraly, E. J. Knol, A. N. Rudenko, M. I. Katsnelson and A. A. Khajetoorians, "Orbital memory from individual Fe atoms on black phosphorus," *Phys. Rev. Res.,* vol. 4, p. 033047, July 2022.

[71] R. Rejali, D. Coffey, J. Gobeil, J. W. González, F. Delgado and A. F. Otte, "Complete reversal of the atomic unquenched orbital moment by a single electron," *npj Quantum Mater.,* vol. 5, p. 60, August 2020.

[72] T. Sauter, W. Neuhauser, R. Blatt and P. E. Toschek, "Observation of Quantum Jumps," *Phys. Rev. Lett.,* vol. 57, p. 1696–1698, October 1986.

[73] A. Singha, P. Willke, T. Bilgeri, X. Zhang, H. Brune, F. Donati, A. J. Heinrich and T. Choi, "Engineering atomic-scale magnetic fields by dysprosium single atom magnets," *Nat. Commun.,* vol. 12, no. 4179, July 2021.

[74] A. Spinelli, M. P. Rebergen, Otte and F. A, "Atomically crafted spin lattices as model systems for quantum magnetism," *J. Phys.:Condens. Matter,* vol. 27, p. 243203, June 2015.

[75] W. Steurer, J. Repp, L. Gross, I. Scivetti, M. Persson and G. Meyer, "Manipulation of the Charge State of Single Au Atoms on Insulating Multilayer Films," *Phys. Rev. Lett.,* vol. 114, p. 036801, January 2015.

[76] K. Vaxevani, J. Li, S. Trivini, J. Ortuzar, D. Longo, D. Wang and J. I. Pascual, "Extending the Spin Excitation Lifetime of a Magnetic Molecule on a Proximitized Superconductor," *Nano Lett.,* vol. 22, p. 6075–6082, July 2022.

[77] R. Wiesendanger, H.-J. Güntherodt, G. Güntherodt, R. J. Gambino and R. Ruf, "Observation of vacuum tunneling of spin-polarized electrons with the scanning tunneling microscope," *Phys. Rev. Lett.,* vol. 65, p. 247–250, July 1990.

[78] R. Wiesendanger, "Spin mapping at the nanoscale and atomic scale," *Rev. Mod. Phys.,* vol. 81, p. 1495–1550, November 2009.

[79] S. Yan, D.-J. Choi, J. A. J. Burgess, S. Rolf-Pissarczyk and S. Loth, "Control of quantum magnets by atomic exchange bias," *Nat. Nanotechnol.,* vol. 10, p. 40–45, December 2014.

[80] J. L. Zhang, J. Q. Zhong, J. D. Lin, W. P. Hu, K. Wu, G. Q. Xu, A. T. S. Wee and W. Chen, "Towards single molecule switches," *Chem. Soc. Rev.,* vol. 44, p. 2998–3022, 2015.

Figures

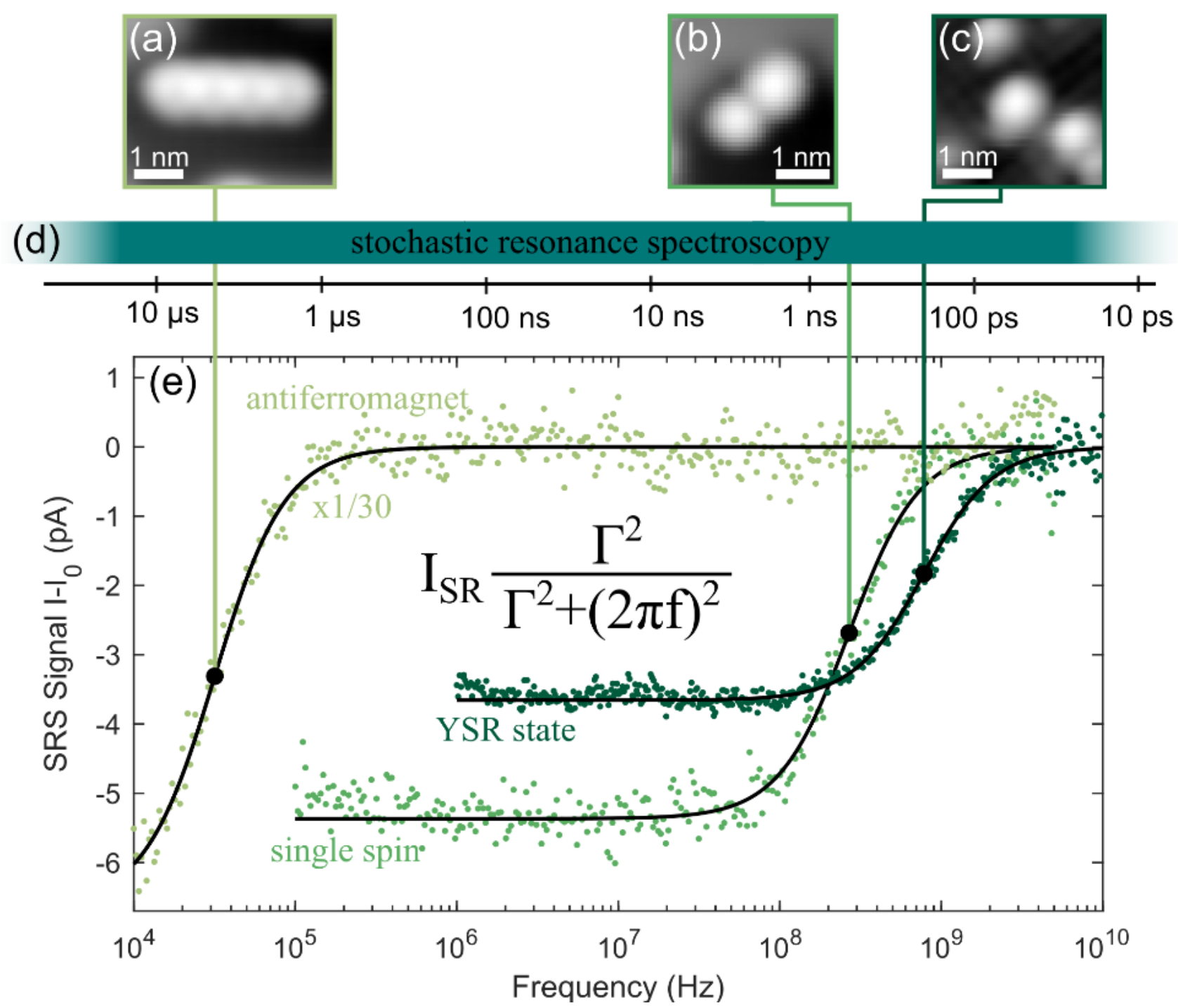

**Figure 1 Stochastic resonance spectroscopy (SRS).** (**a-c**) STM topographies of (**a**) nano-antiferromagnet on a $Cu_2N$ decoupling layer on Cu(100), (**b**) individual Fe atom on the same surface, and (**c**) an Fe atom on the superconducting V-(1x5)O surface. Tunnel junction setpoints (**a**,**b**) 200 pA at 100 mV, (**c**) 108 pA at 50 mV. Scale bars indicate 1 nm length. (**d**) Illustration of the time scales accessible to measurements using SRS (teal ribbon). Lifetimes of the three systems in (a)-(c) are marked with vertical lines. (**e**) Stochastic resonance spectroscopy of the nano-antiferromagnet (light green), single spin of the Fe atom on $Cu_2N$ (green) and Yu-Shiba-Rusinov state of the Fe atom on V (dark green), showing the variation of the tunnel current as function of microwave frequency. The measurement on the antiferromagnetic chain shows a step in current of $199 \pm 2$ pA amplitude, centered at $f_{AFM} = 31.7\ kHz$ (marked by a black dot). The measurement on the single atom shows a step in current of $5.37 \pm 0.05\ pA$ amplitude, centered at $f_{Fe} = 269\ MHz$ (marked by a black dot) and the one of the YSR state shows a step of $3.66 \pm 0.02\ pA$ amplitude, centered at $f_{YSR} = 781\ MHz$. Fits to the data using eq. 1 (black lines) yield transition rates of $\Gamma_{\text{AFM}} = 199 \pm 5 \cdot 10^3/s$ for the nano-antiferromagnet, $\Gamma_{Fe} = 1.69 \pm 0.05 \cdot 10^9/s$ for the single spin and $\Gamma_{\text{YSR}} = (4.91 \pm 0.06) \cdot 10^9/\text{s}$ for the YSR state. The setpoints are: 4 nA at 10 mV with microwave amplitude $V_{ac} = 1.4\ mV$ and offset voltage $V_{dc} = 4\ mV$ (antiferromagnet), 4 nA at 10 mV with $V_{ac} = 2.8\ mV$ and $V_{dc} = 4\ mV$ (single spin), and 1 nA at 10 mV with $V_{ac} = 140\ \mu V$ and $V_{dc} = 640\ \mu V$ (YSR state). Magnetic field 2 $T$, temperature 2.4 $K$ (antiferromagnet and single spin) and 0 $T$, 50 $mK$ (YSR state).

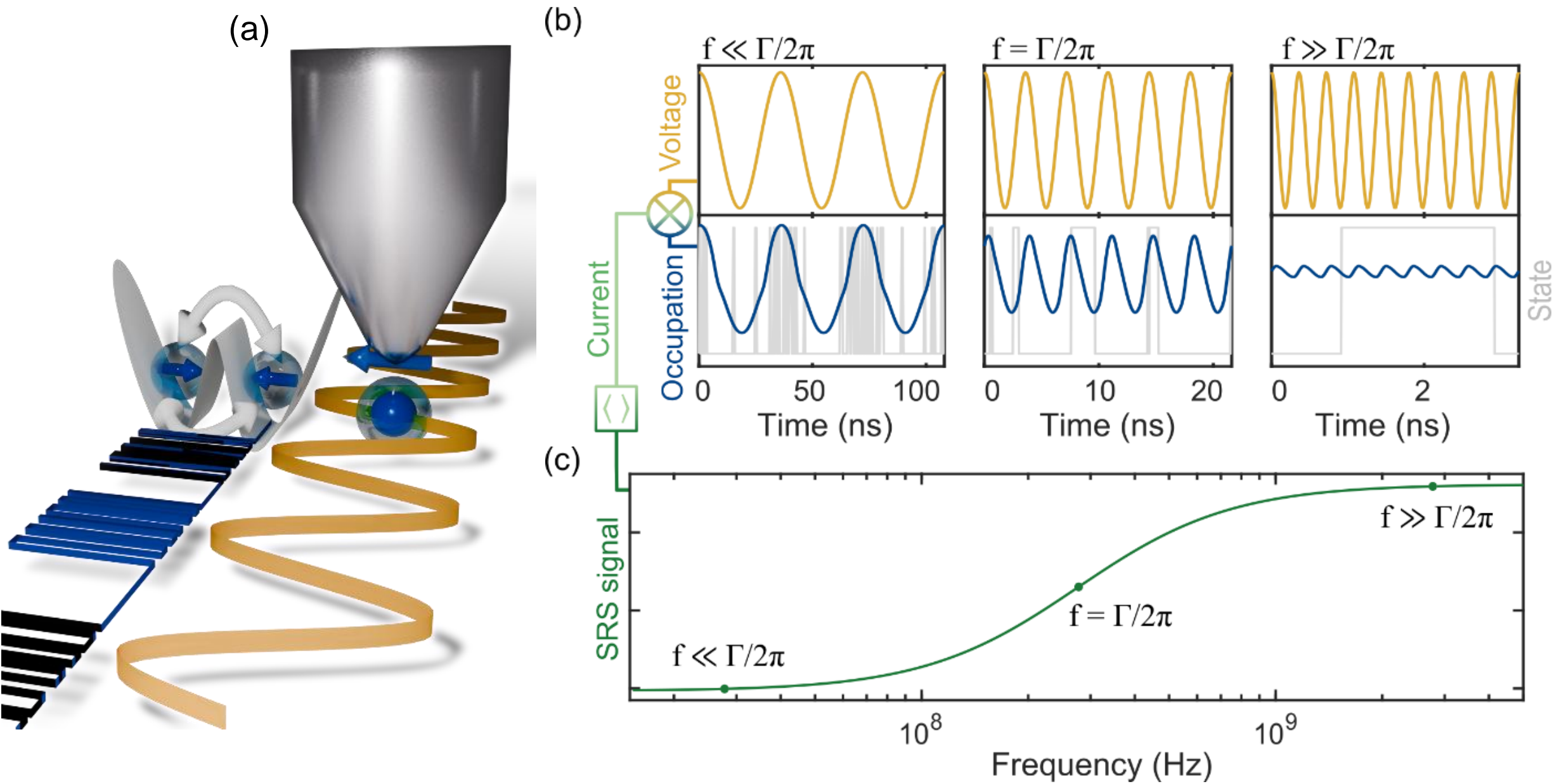

**Figure 2 Detection mechanism of stochastic resonance spectroscopy.** (**a**) Sketch of the effect of the microwave drive (yellow ribbon) on the spin switching (blue ribbon) at microwave frequencies $f \ll \Gamma/2\pi$. The drive, applied to the atom (blue sphere) using a spin-polarized tip (indicated by blue arrow at the tip apex), facilitates transitions (white arrows) between the spin states (blue arrows), that are separated by an energy barrier. (**b**) The voltage of the microwave modulation (yellow) mixes with the spin system's occupation probability (blue) and representative state occupation (grey). The time-dependent occupation probability changes with microwave frequency $f$ (left to right: $28\,MHz\!: f \ll \Gamma/2\pi$, $278\,MHz\!: f = \Gamma/2\pi$ and $2.7\,GHz\!:: f \gg \Gamma/2\pi$). (**c**) Resulting SRS signal (see eq. 1 in the main text), given by low-passed detection of the tunnel current.

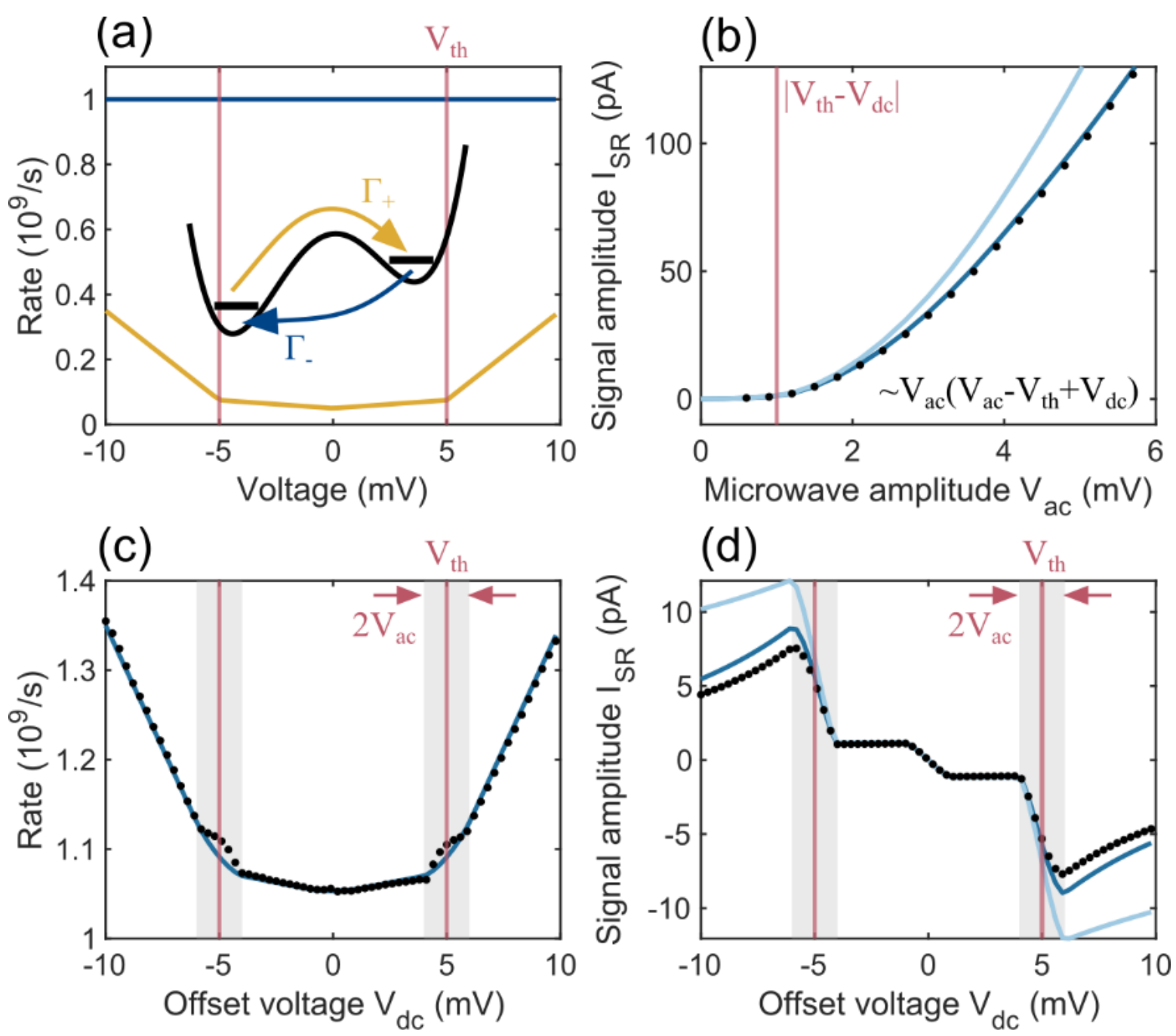

**Figure 3 Factors determining the amplitude of the SRS signal.** (**a**) Excitation, $\Gamma_+$, (yellow) and relaxation rates, $\Gamma_-$, (blue), of a two-state system with an energy threshold (red) at $V_{th} = \pm 5\, meV$. (**b**) SRS signal amplitude, $I_{SR}$, as function of microwave amplitude, $V_{ac}$, for the two-state system of (a) obtained by fitting eq. 1 (black dots) to a numerical model compared to the analytical model of SRS up to 1$^{st}$ order (light blue curve) and 2$^{nd}$ order (dark blue). Above the minimal $V_{ac}$ (red line), where the voltage surpasses $V_{th}$ for $V_{dc} = -4\, mV$, the SRS signal is proportional to $V_{ac}(V_{ac} - V_{th} + V_{dc})$. (**c**) Transition rates extracted from the SRS signal by fitting eq. 1 (black dots) compared to the total transition rate $\Gamma = \langle \Gamma_+(t) + \Gamma_-(t) \rangle$ of (a) (dark blue). The fitted rates deviate from the analytical rates by up to 2% in a voltage interval of $2V_{ac}$ centered around $\pm V_{th}$ where the excitation rate changes non-linearly (grey area). (**d**) SRS signal amplitude as function of offset voltage, $V_{dc}$ , obtained by fitting eq. 1 (black dots) compared to the analytical model of SRS up to 1$^{st}$ order (light blue) and 2$^{nd}$ order (dark blue). The signal amplitude changes approximately linearly within a voltage interval of $2V_{ac}$ centered around $\pm V_{th}$ (grey area) and $V_{dc} = 0\, meV$.

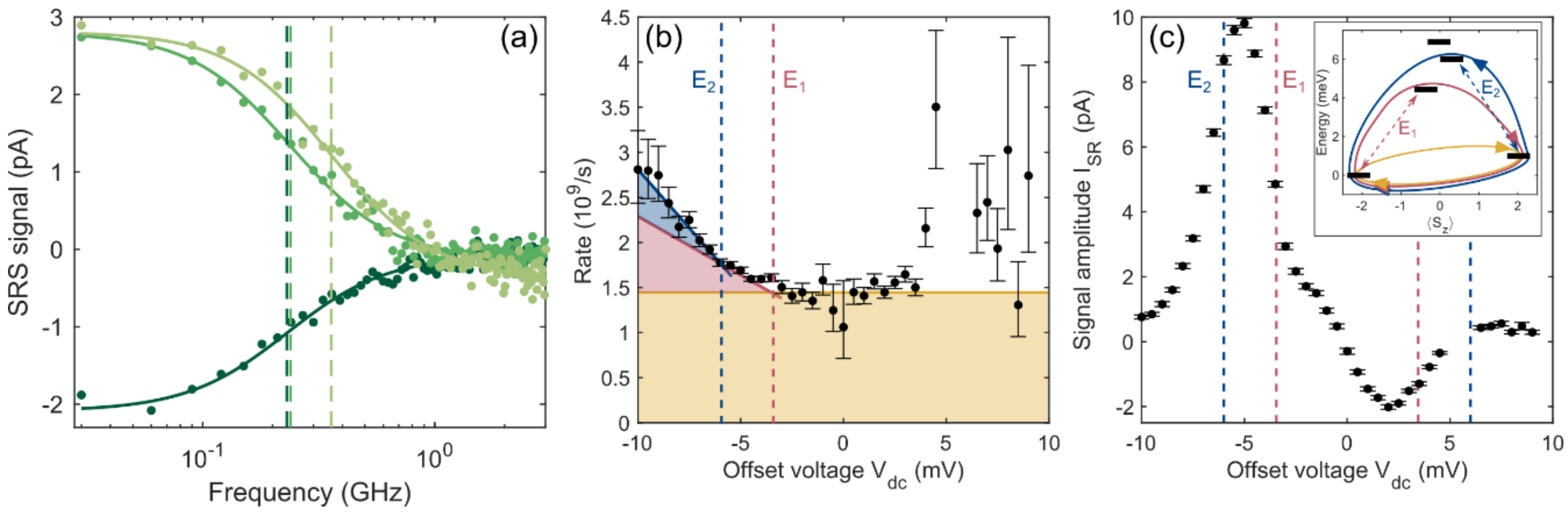

**Figure 4 Driven dynamics of a single Fe atom on $Cu_2N$.** (**a**) SRS measurements at +2 mV (dark green dots), −2 mV (green dots) and −6 mV (light green dots). Solid lines are fits using eq. 1 with their respective $f = \frac{\Gamma}{2\pi}$ frequencies indicated by dashed vertical lines. (**b**) Transition rates (black dots) extracted from SRS measurements at different $V_{dc}$ analogous to (a). Dashed lines indicate energy thresholds $E_1$ (red) and $E_2$ (blue) of the Fe spin. Shaded regions indicate transition rates caused by direct transitions below $E_1$ (yellow), transitions over $E_1$ (red) and transitions over $E_2$ (blue). Positive offset voltage side not shaded because of large fit uncertainties. (**c**) SRS signal amplitude (black dots) extracted from same measurements as in (b). Energy thresholds $E_1$, $E_2$ indicted by dashed lines. Inset: Fe atom spin states (black bars) as function of their $\langle S_z \rangle$ expectation values and energy, showing the transition pathways for direct transitions (yellow), transitions over $E_1$ (red) and over $E_2$ (red). Note that transversal anisotropy mixes multiple spin-states and thus causes non-integer $\langle S_z \rangle$ values [40]. Error bars in (b,c) indicate least-squares fit uncertainties. Measurement setpoint 1 nA at 10 mV, microwave amplitude $V_{ac} = 1.3\ mV$, magnetic field 2 T, temperature 0.5 K.

**Supplemental Material:**

# Quantifying Atomic-Scale Transition Rates Using Stochastic Resonance Spectroscopy

Nicolaj Betz[1,2], Gregory McMurtrie[1], Max Hänze[1,3], Vivek Krishnakumar Rajathilakam[1], Laëtitia Farinacci[1,4], Susan N. Coppersmith[5], Susanne Baumann[1] and Sebastian Loth[1,2,3]

[1] University of Stuttgart, Institute for Functional Matter and Quantum Technologies, 70569 Stuttgart, Germany.
[2] Center for Integrated Quantum Science and Technology (IQST), University of Stuttgart, 70569 Stuttgart, Germany.
[3] Max Planck Institute for Solid State Research, 70569 Stuttgart, Germany.
[4] Carl-Zeiss-Stiftung Center for Quantum Photonics Jena – Stuttgart – Ulm, Germany.
[5] School of Physics, University of New South Wales, Sydney, Australia.

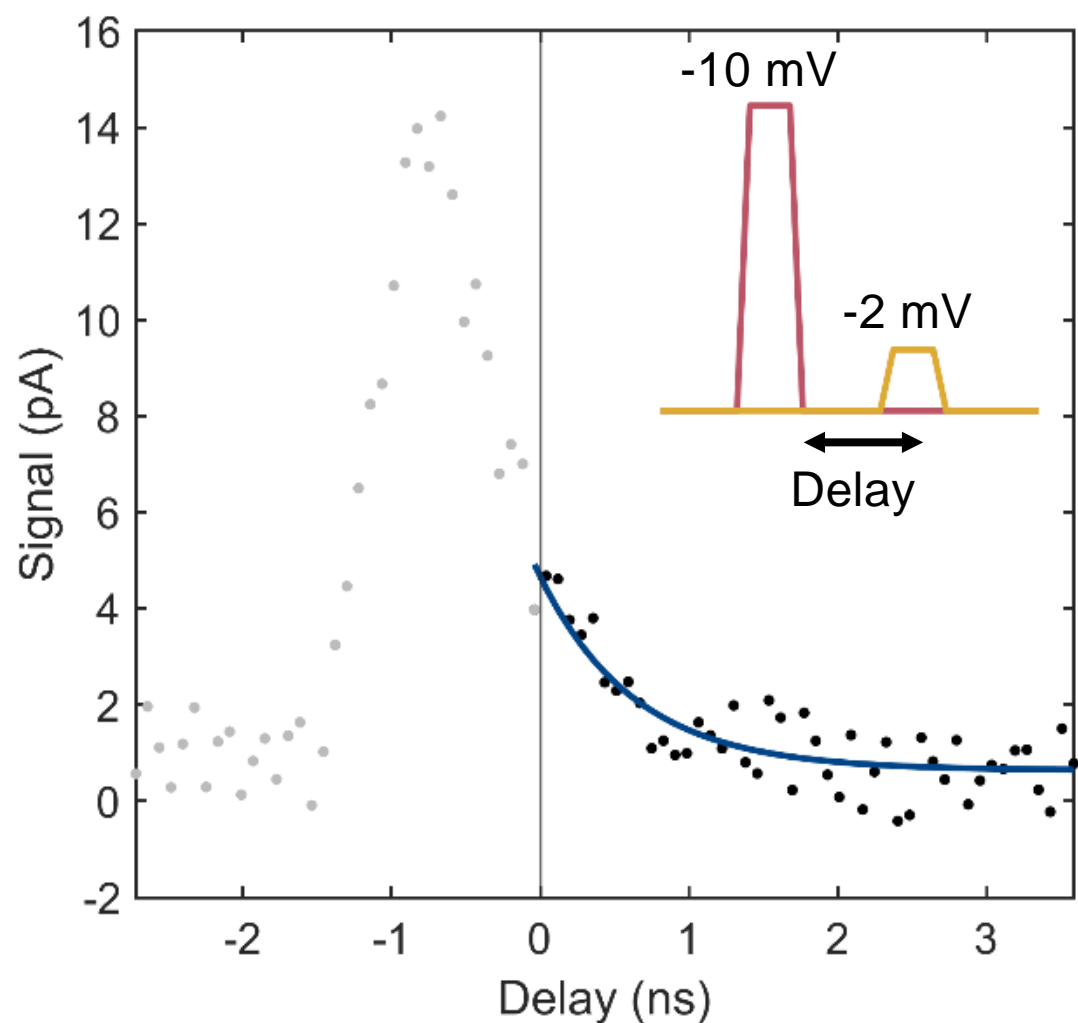

***Supplementary Figure S1 Electronic pump-probe measurement of a single Fe atom on $Cu_2N$.*** *In this experiment, two 429 ps voltage pulses (pump: -10 mV, probe: -2 mV) with rise times of 154 ps and varying delay are applied to the tunnel junction. The resulting current signal decays as the excited spin population relaxes [27]. The exponential fit to the signal decay (dark blue) results in a lifetime of* $T_1 = 0.6 \pm 0.1\ ns$. *The spin-polarized tip in this measurement is the same as the tip during the measurements on the Fe atom in Fig. 1 of the main text. The measurement has been performed using an active state setpoint voltage and current of* $V_{set} = 10\ mV$, $I_{set} = 5\ nA$, *in a magnetic field of* $B = 2\ T$ *and at a temperature* $T = 2.4\ K$.

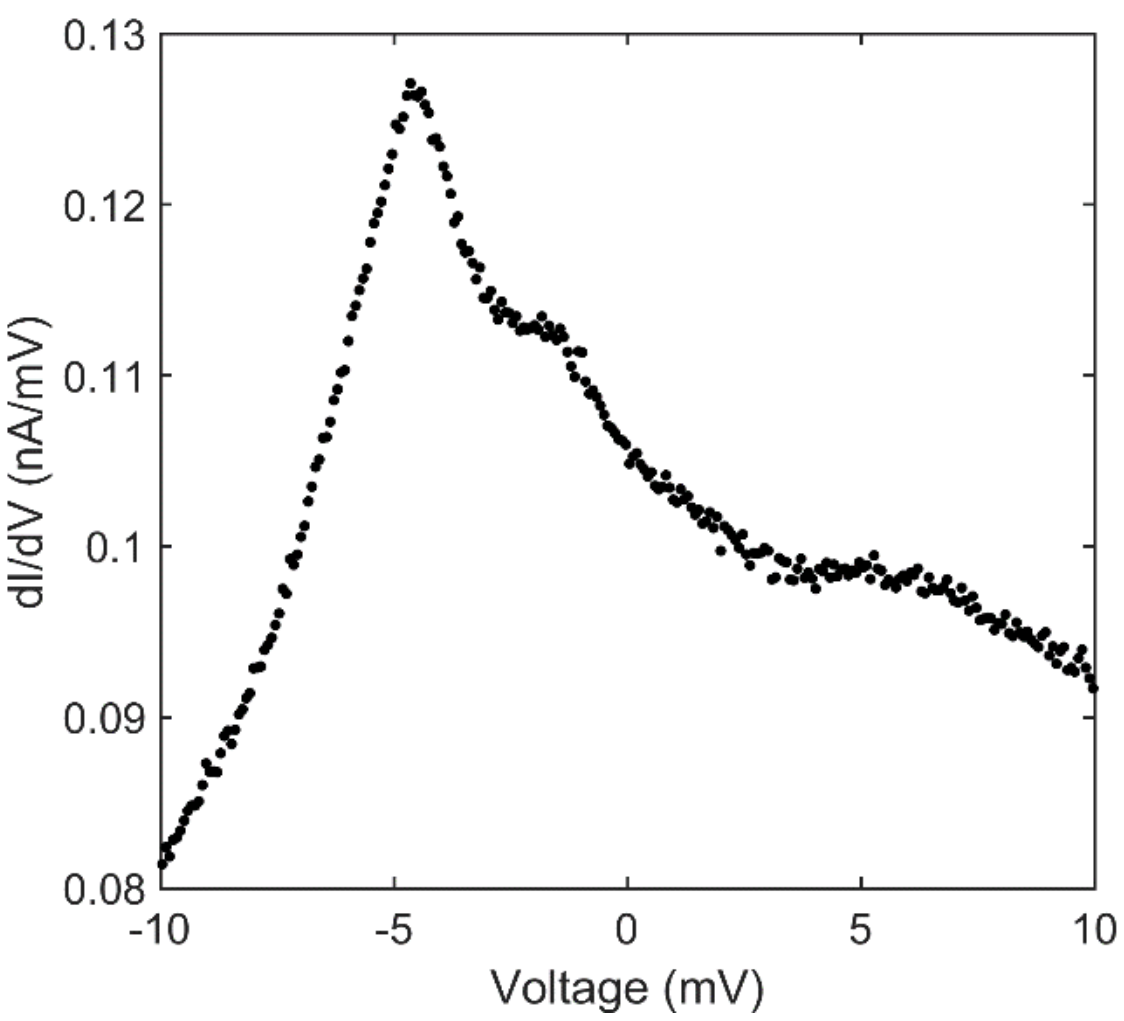

***Supplementary Figure S2 Differential conductivity $dI/dV$ of a single Fe atom on $Cu_2N$, measured with a spin-polarized tip.*** *The spin-polarized tip in this measurement is the same as the tip during the measurements on the Fe atom in Fig. 1 of the main text. The measurement has been performed using an active state setpoint voltage and current of* $V_{set} = 10\ mV$, $I_{set} = 1\ nA$, *in a magnetic field of* $B = 2\ T$ *and at a temperature* $T = 2.4\ K$.

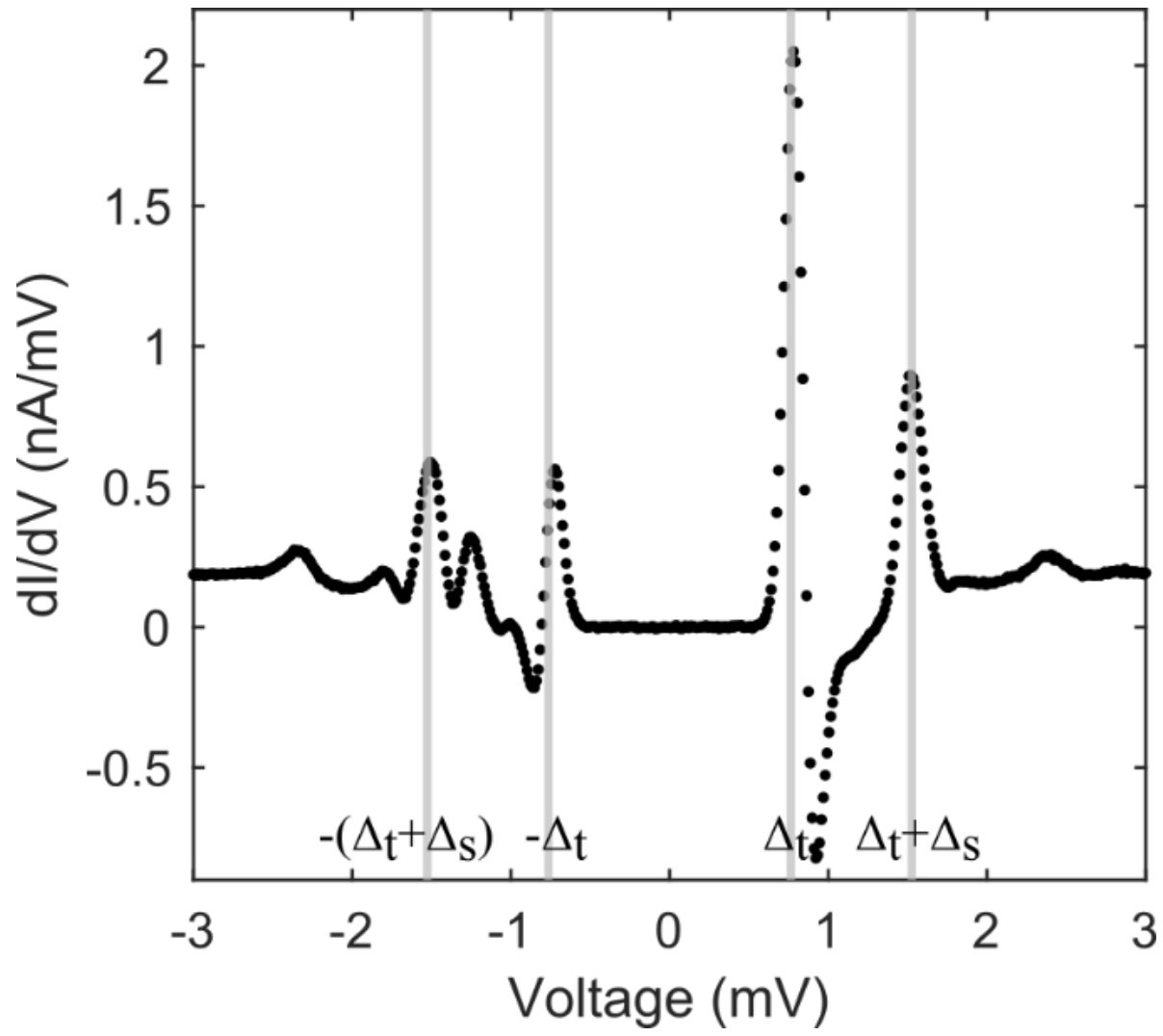

***Supplementary Figure S3 Differential conductance (dI/dV) of the Yu-Shiba-Rusinov state in Fig. 1.*** *The dI/dV is recorded with a superconducting tip on an Fe atom on V(100)(1×5) (black dots) shows a strong peak-dip feature close to the excitation thresholds of the tip $\Delta_t$ and multiple weaker peaks between $\Delta_t$ and $\Delta_s + \Delta_t$ (marked in gray). The measurement has been performed using an active state setpoint voltage and current of $V_{set} = 10\ mV$, $I_{set} = 2\ nA$ and at a temperature $T = 50\ mK$.*

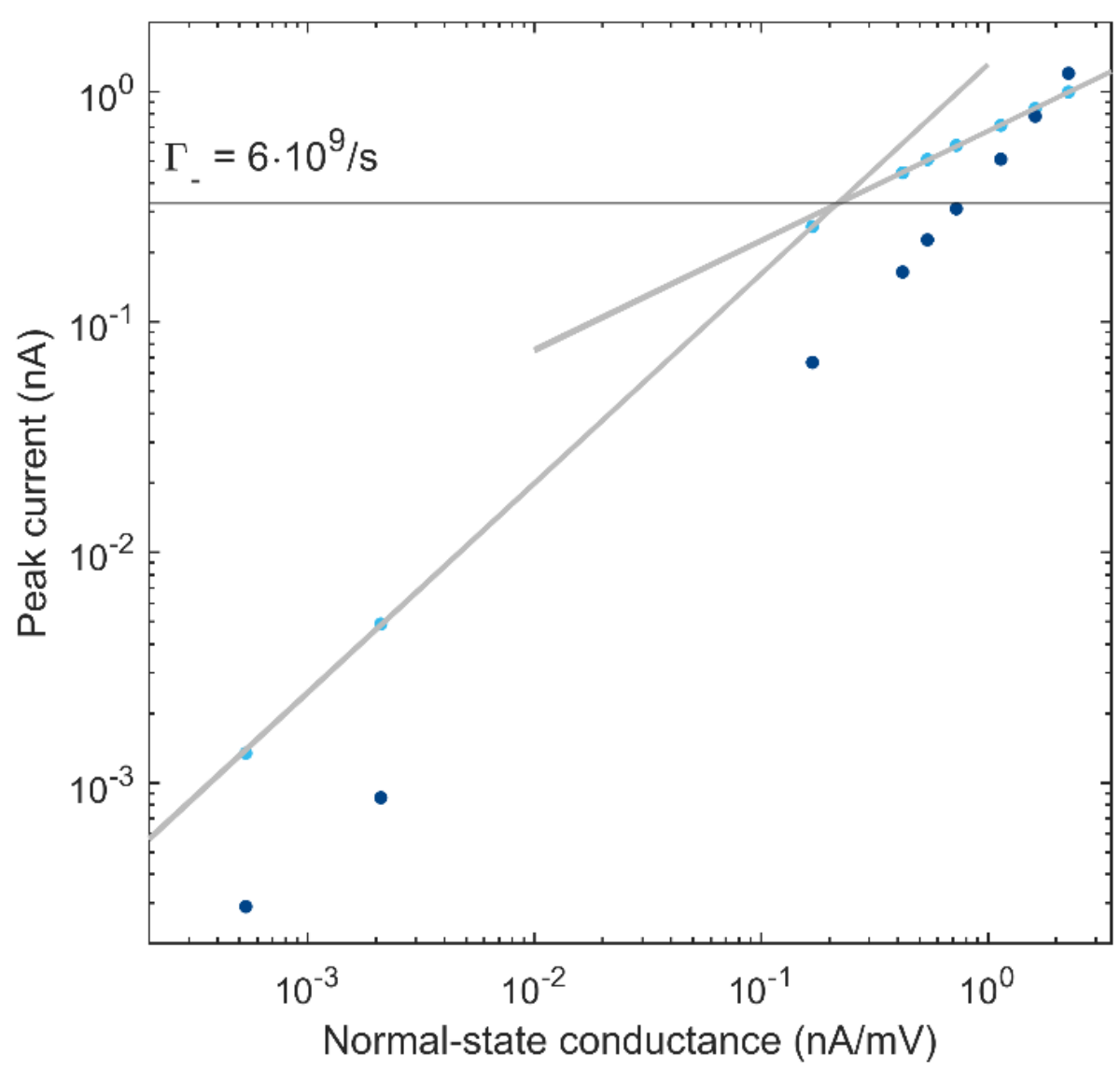

***Supplementary Figure S4 Maximum tunneling current through a Yu-Shiba-Rusinov state on V(100)***. *The tunneling current through the Yu-Shiba-Rusinov state of an Fe atom on V(100) as used in Fig. 1 of the main text is displayed as a function of junction conductance. At positive voltages (light blue data points) the current displays a linear and a sub-linear regime, fitted individually using a power law (grey lines). While the current at negative voltages (dark blue) does not display a transition between linear and sub-linear regimes. For positive voltages, the current at the intersection between the linear and sub-linear regime (black line) estimates the relaxation rate of the system to $\Gamma_{-} \approx 6 \cdot 10^9/s$. The normal-state conductance is obtained through the current at 3 mV. All measurements are taken at T = 50 mK.*

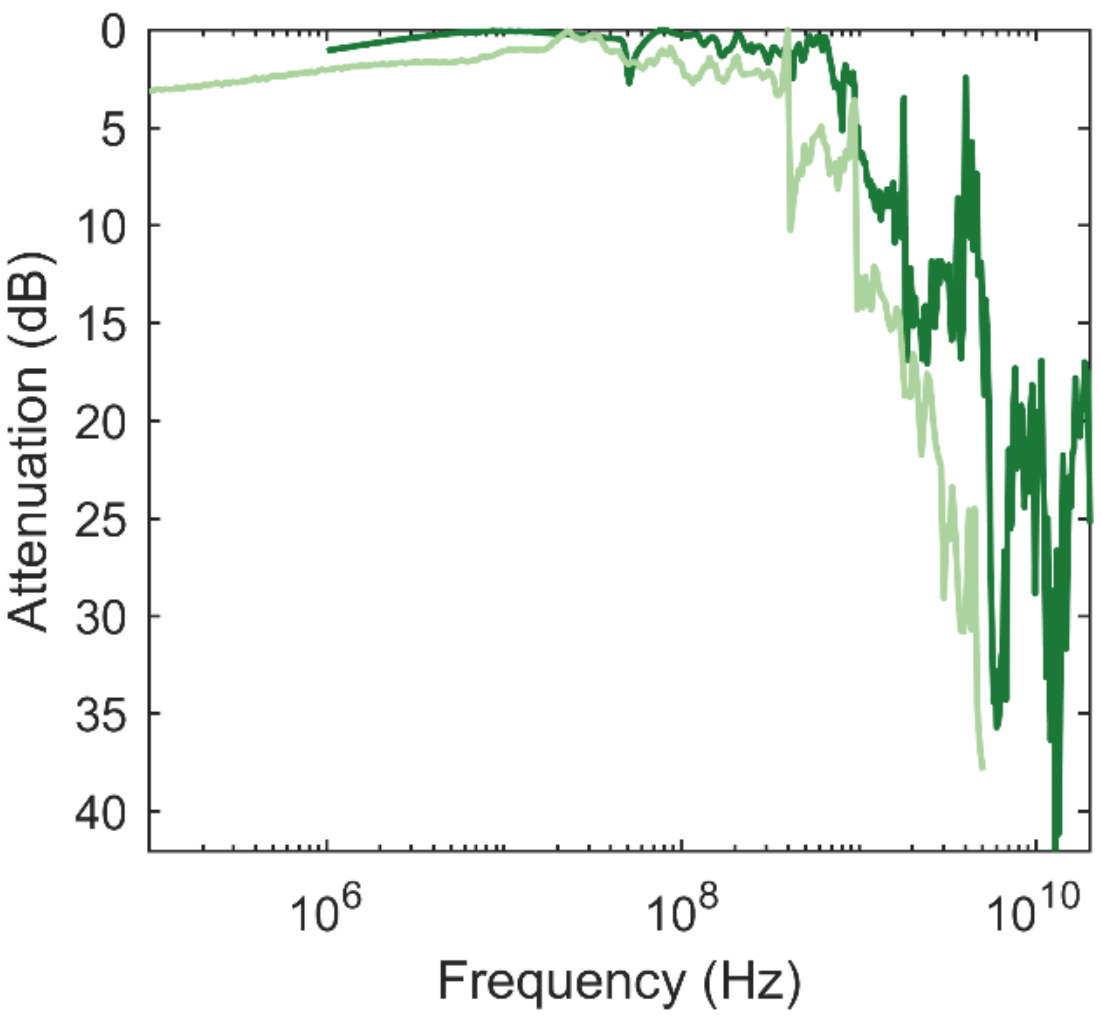

***Supplementary Figure S5 Attenuation of microwave signals in the experimental setups.*** *Characterization of the attenuation of the Unisoku USM-1300 (light green) has been measured on an Fe atom adsorbed on a Cu site on* $Cu_2N$ *with anisotropy axis perpendicular to the external magnetic field* $B = 2\,T$*. The characterization has been performed with an amplitude target of* $V_{ac} = 2.8\,mV$ *and an offset voltage of* $V_{dc} = -4\,mV$ *(*$T = 2.4\,K$*,* $I_{set} = 4\,nA$*,* $V_{set} = 10\,mV$*). Characterization of the attenuation of the Unisoku USM-1600 (dark green) has been performed on the bare V(100)(1×5) surface, an amplitude target of* $V_{ac} = 120\,\mu V$ *and an offset voltage of* $V_{dc} = 1.4\,mV$ *(T=50 mK,* $I_{set} = 1\,nA$*,* $V_{set} = 10\,mV$*).*

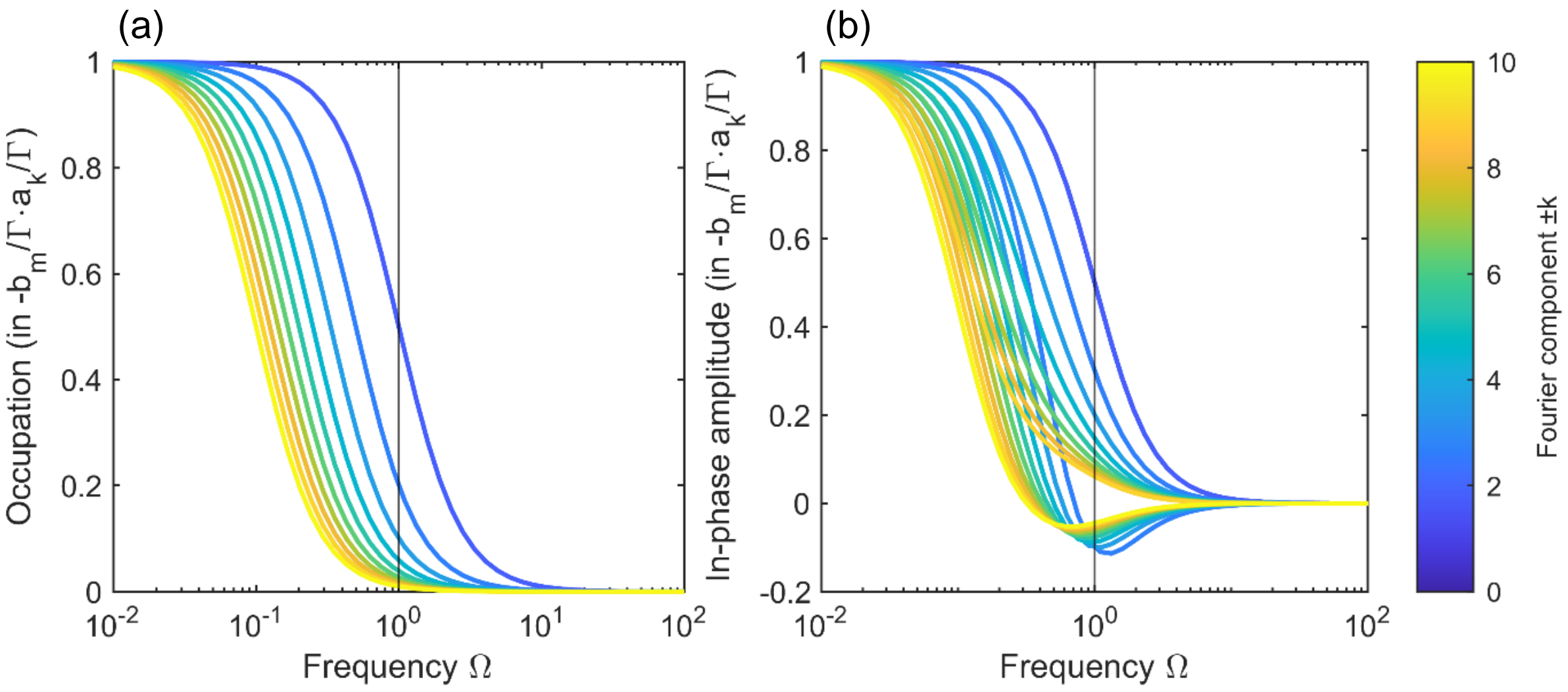

***Supplementary Figure S6 Frequency-dependence of 2nd order terms*** *The figure shows the* $j = 2$*-terms of eq.* [19] *(see section S1.2 in supplemental material) that (a) contribute to the average occupation* $(m + k = 0)$ *and (b) contribute to the in-phase occupation amplitude oscillating at* $\omega$*,* $(m + k = \pm 1)$*. Each curve is normalized by the respective Fourier components* $-b_m/\Gamma\ \ a_k/\Gamma$ *, encoded in the color scale.*

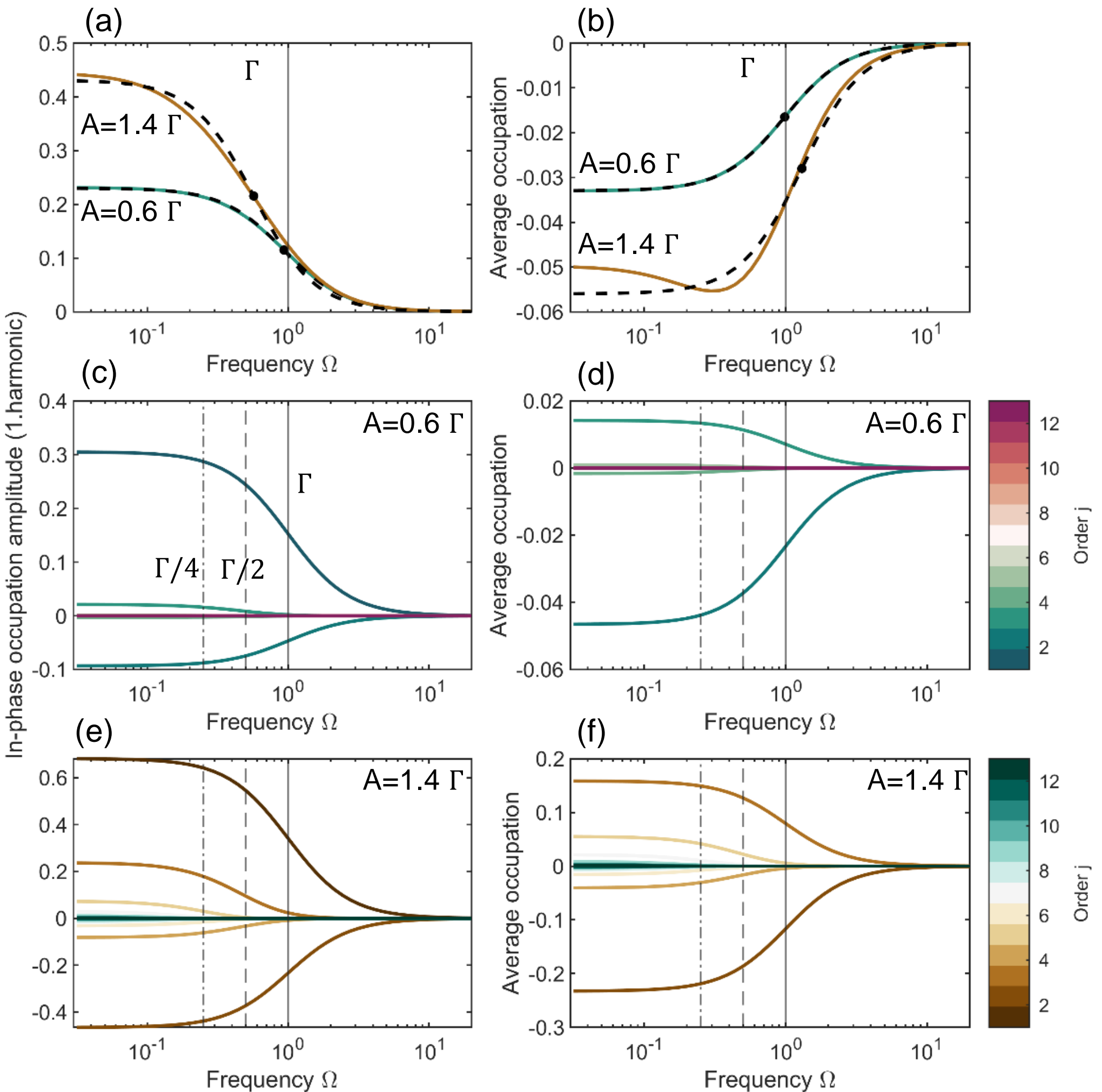

***Supplementary Figure S7 Contribution of different orders j to the average occupation and the in-phase occupation amplitude, oscillating at ω.*** *(**a, b**) Total in-phase occupation amplitude and average occupation at modulation strengths $A = 0.6\,\Gamma$ (purple) and $1.4\,\Gamma$ (yellow), plotted together with fits, using eq. 1 of the main text (dashed lines). The extracted rates are marked with a dot. The rates due to a modulation strength of $A = 0.6\,\Gamma$ are shown in supplementary Fig. S8a, while those for the larger modulation amplitude would increase accordingly. (**c, e**) Contribution of terms of order j up to $j_{max} = 13$ to the in-phase occupation amplitude for the two different modulation strength $A = 0.6\,\Gamma$ (**c**) and $1.4\,\Gamma$ (**e**). (**d, f**) Contribution of terms of order j up to $j_{max} = 13$ to the average occupation for the respective modulation strength. Respective orders are displayed according to the color scale on the right. Gray lines mark $\Gamma$ (solid lines), $\Gamma/2$ (dashed lines) and $\Gamma/4$ (dashed-dotted lines). Overall, this figure shows how larger modulation strengths promote higher order contributions that shift the total signal toward lower frequencies and cause distortions in the measured SRS signal.*

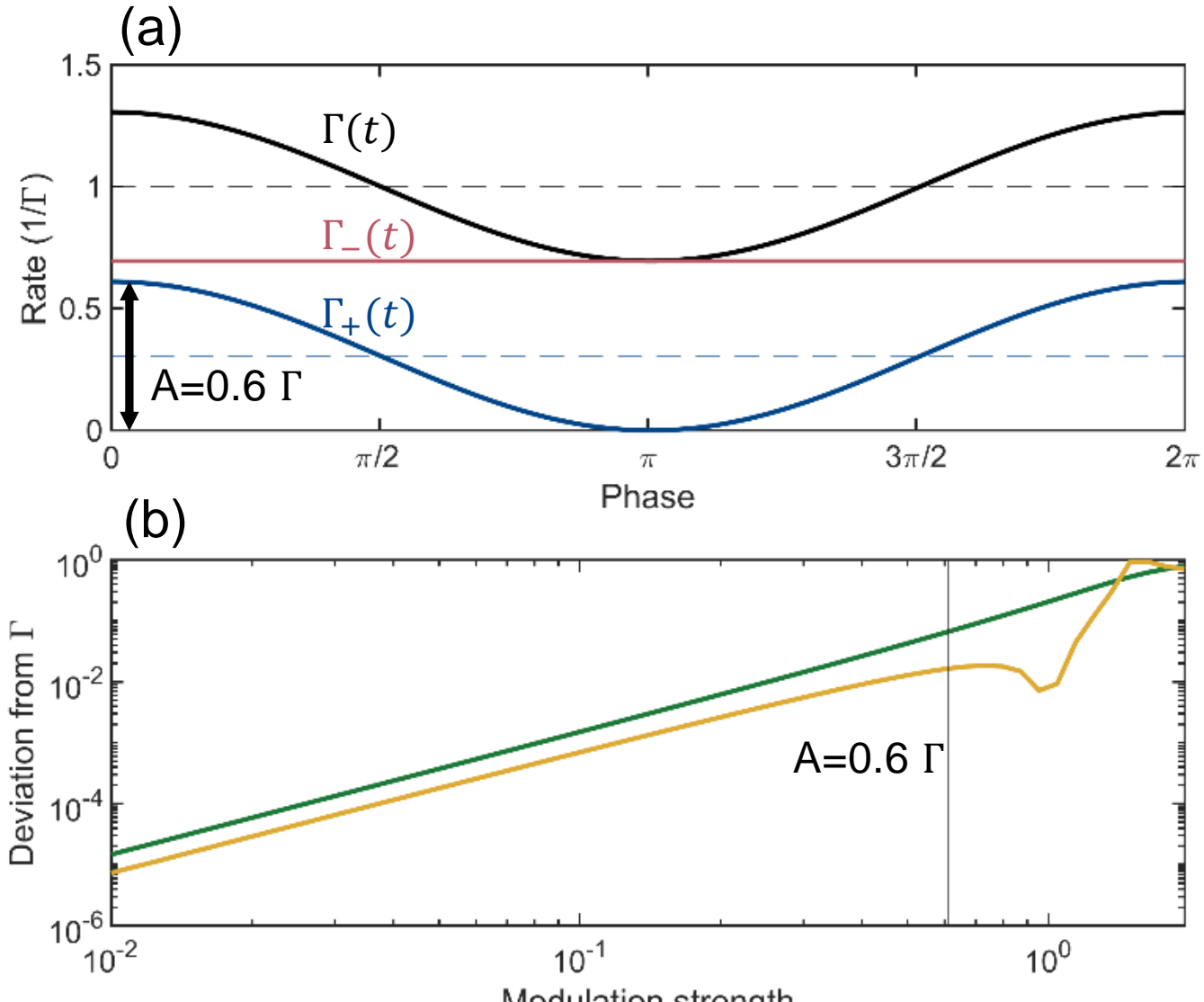

***Supplementary Figure S8 Effect of large variations in the total transition rate*** *(**a**) Excitation, relaxation and total rates for modulation strength of $A = 0.6\,\Gamma$ in units of $\Gamma$. The excitation rate $\Gamma_+(t) = \frac{A}{2}(1 + cos(\omega t))$ (blue) modulates sinusoidally from 0 to A while the relaxation rate $\Gamma_- = \Gamma - \langle\Gamma_+\rangle$ (red) is constant. The resulting total rate $\Gamma(t) = \Gamma_+(t) + \Gamma_-$ is displayed in black. Dashed lines mark the average of their respective rates. (**b**) Such large modulation strength lead to deviations in the extracted rate. These deviation from the actual rate $\Gamma$ are displayed here: increasing modulation strength leads to larger deviations from the rate, typically shifting the extracted rate to smaller $\Gamma$ (as shown in supplementary Fig. S7a). The plotted deviations are obtained by fitting the theoretical signal, due to the average occupation probability (yellow) and due to homodyne mixing (green), with eq. 1 of the main text. These deviations can be understood as the error in $\Gamma$ when using eq. 1, due to higher-order contributions. The theoretical signal is calculated including all contributions up to order $j = 13$. Modulation strengths up to $A = 0.6\,\Gamma$ (gray line) result in deviations of less than 5% of the actual rate.*

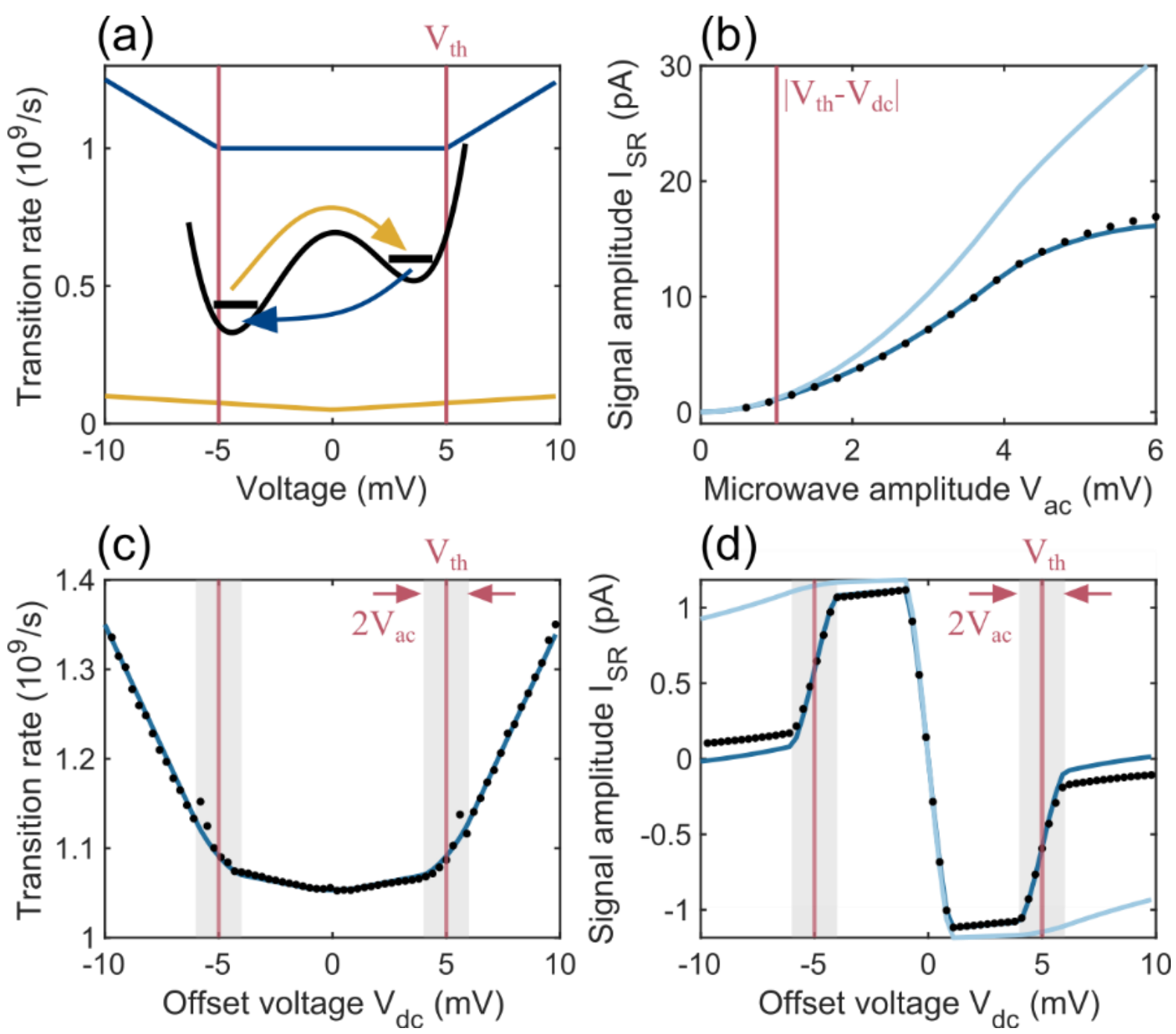

***Supplementary Figure S9 SRS signal with threshold in relaxation rate.*** *(**a**) Excitation $\Gamma_{ex}$ (yellow) and relaxation rate $\Gamma_{rel}$ (blue) of a two-state system (illustrated in inset) with a voltage threshold at $V_{th} = \pm 5\ mV$ (red). (**b**) SRS signal amplitude as function of microwave amplitude and a voltage offset of $V_{dc} = -4\ mV$. (**c**,**d**) Total transition rate $\Gamma$ and SRS signal amplitude as function of offset voltage and $V_{ac} = 1\ mV$. Black dots are obtained though numerical simulation of the SRS measurement and subsequent fitting using eq. 1, dark- and light-blue lines in (**b**,**d**) are calculated according to the analytical theory up to 1st and 2nd order, respectively, and the blue line in (**c**) is the total transition rate $\Gamma = \langle \Gamma_{ex}(t) + \Gamma_{rel}(t) \rangle$ (from (a)). All calculations assume a junction conductivity of $1\ nA/mV$ when the system is in the ground state and a conductivity difference of $\Delta\sigma = -0.5\ nA/mV$ between the states.*

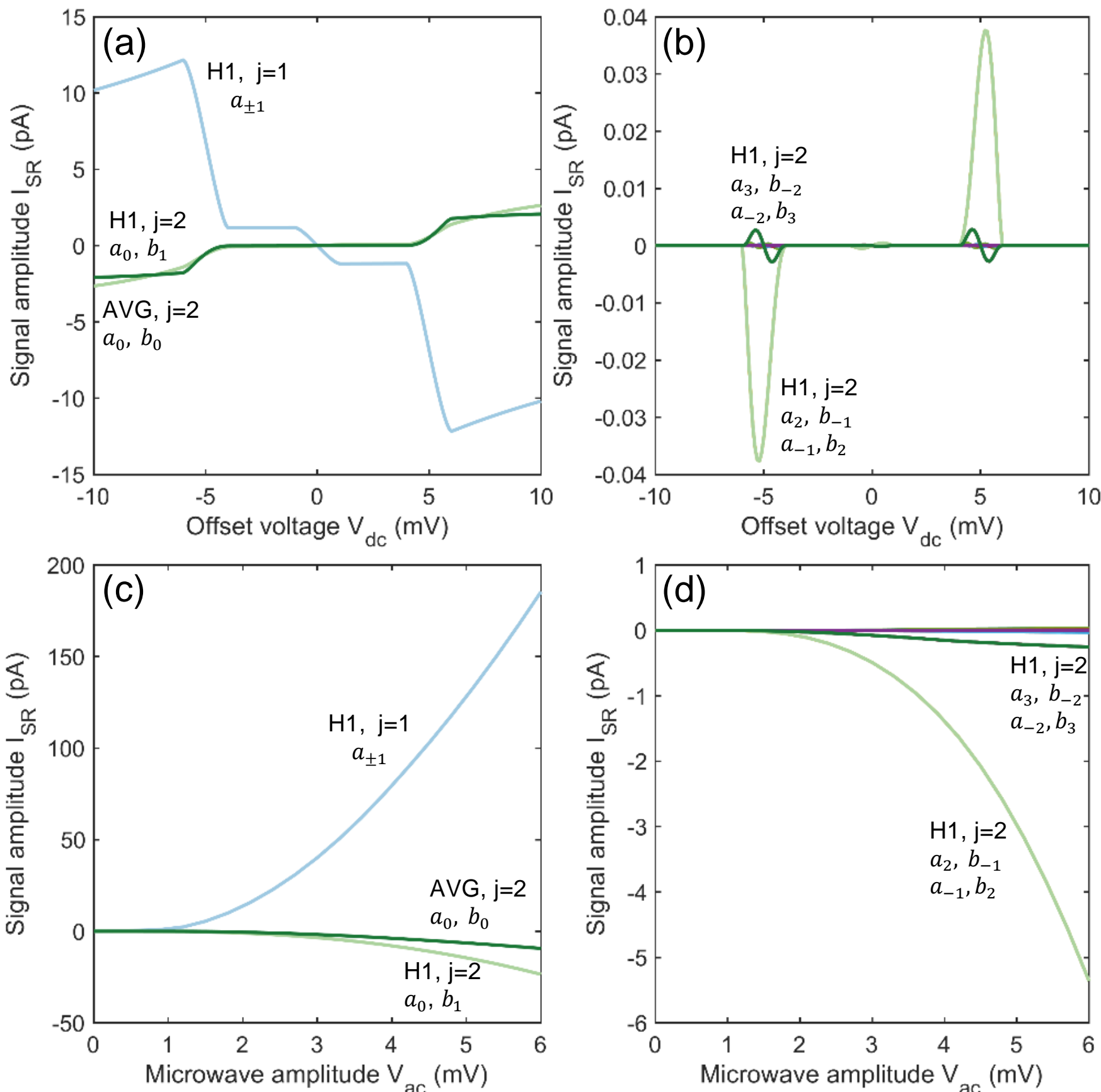

***Supplementary Figure S10 Contributions to the SRS signal amplitude in Fig. 3b and 3d of the main text***. *(**a**,**b**) Contributions to the SRS signal amplitude as a function of offset voltage $V_{dc}$ (**c**,**d**) Contributions to the SRS signal amplitude as a function of microwave drive amplitude $V_{ac}$. The main deviations from the lowest order behavior (H1,j=1) appear around the excitation threshold. The curves are denoted by their order j, the type of their contribution (homodyne signal H1 or average occupation signal AVG) and by their respective Fourier components ($a_k$, $b_m$) in the plots. Contributions with $a_{|k|\geq 2}$, that distort the SRS lineshape are displayed in (**b**,**d**). Those that do not distort the SRS lineshape are displayed in (**a**,**c**). All contributions are calculated according to the analytical theory and use the rates obtained from scattering theory.*

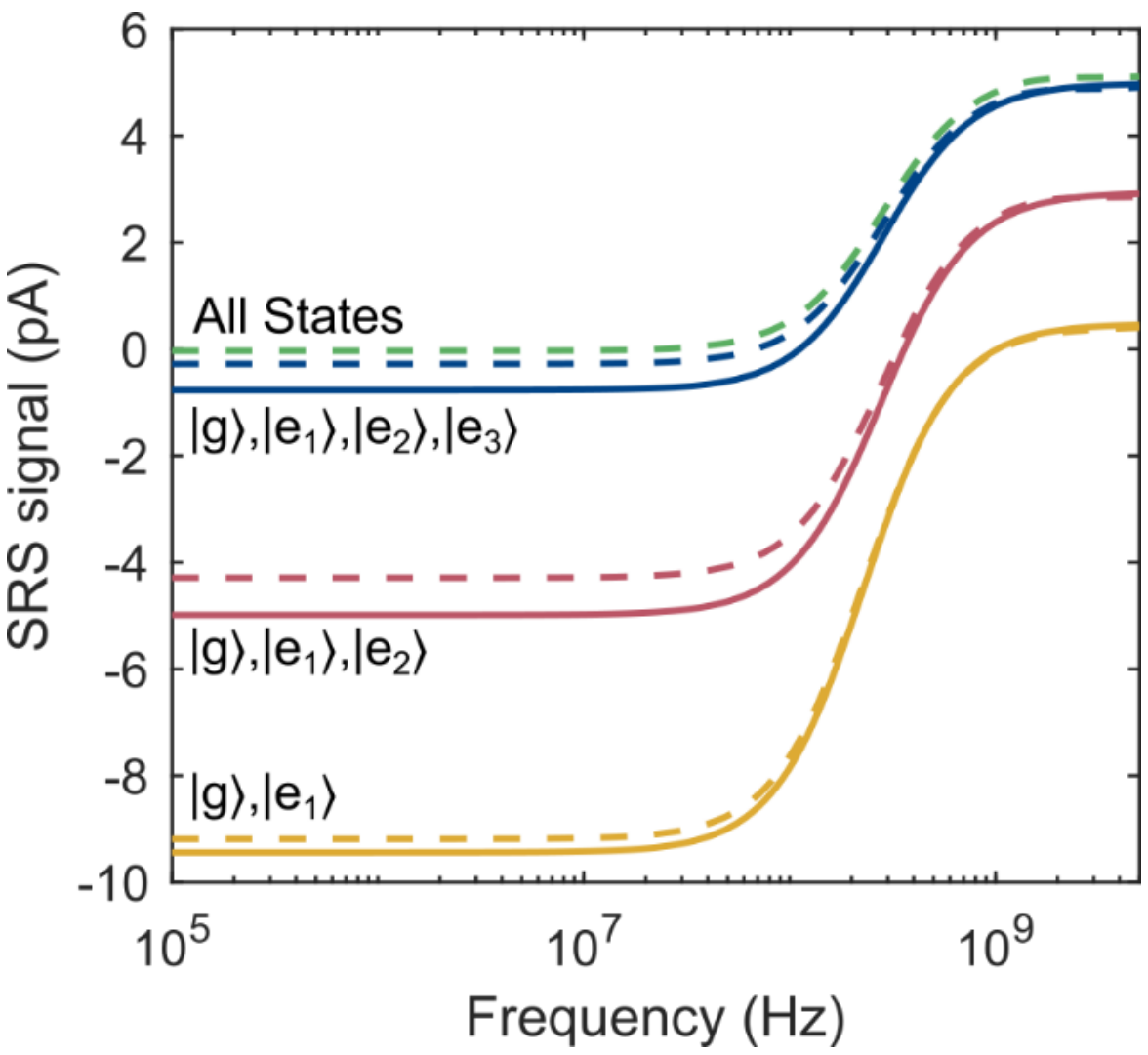

***Supplementary Figure S11 SRS signal upon including different sets of states*** *$|g\rangle, |e_1\rangle$ to $|e_4\rangle$ in the simulation of the Fe atom on $Cu_2N$. The SRS signal is obtained through numerical simulation (dotted lines) as described in supplementary section S3, as well as by directly calculating the theoretical step height and position up to order $j = 2$ (solid lines). Both calculations are based on the voltage dependent transition rates, obtained through scattering theory, as described in supplementary section S4 and the parameters listed in set 1 in supplementary table 1. The parameters have been obtained by fitting the simulation including all states (green) to the measurement displayed in green in Fig. 1 of the main text. Simulations with fewer states mainly shift the offset of the SRS signal. Additionally, excluding states removes transitions through the excluded states and thus reduces the total transition rate. This is reflected by a slight shift of the step positions toward lower frequencies.*

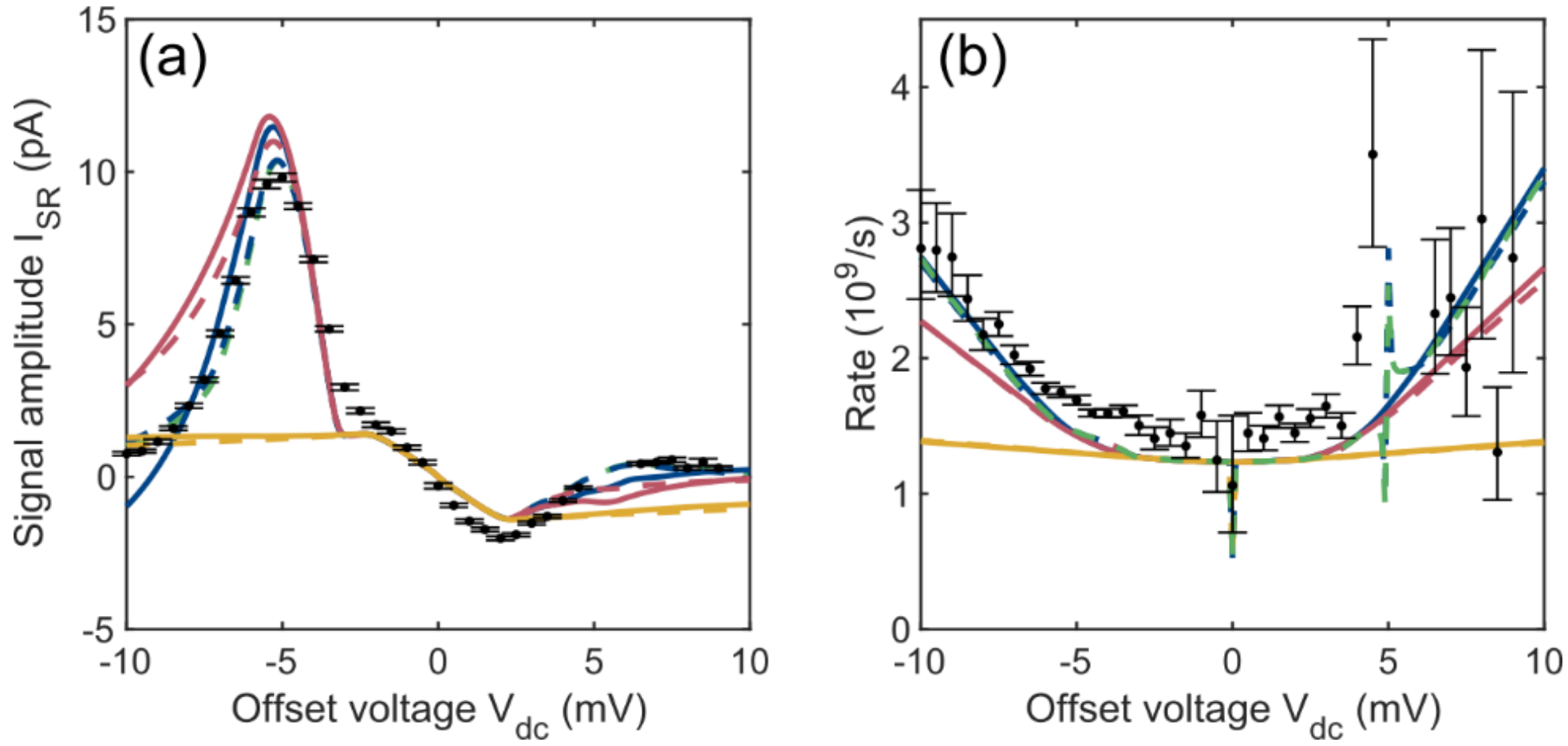

***Supplementary Figure S12 Simulation of the driven dynamics of a single Fe atom on $Cu_2N$*** *displayed in Fig. 4 of the main text. (a) SRS signal amplitude according to the analytical theory up to order $j = 2$ (solid lines) and obtained through numerical simulation (dotted lines), as described in supplementary section S3, fitted with eq. 1 of the main text. (b) Average of the sum of the transition rates $\Gamma = \langle \Gamma_+(t) + \Gamma_-(t) \rangle$ (solid lines) and transition rate obtained through numerical simulation (dotted lines). All calculations are based on the voltage dependent transition rates, obtained through scattering theory, as described in supplementary section S4 and the parameters listed in set 2 in supplementary table 1. As in supplementary Fig. S11, the rates are calculated using only the ground and 1st excited state (yellow), the ground, 1st and 2nd excited states (red), the ground and 1st to 3rd excited states (blue) and all states (green). Consistent with the results in section VI in the main text, up to offset voltages of $|V_{dc} \pm V_{ac}| < 4\,mV$ all simulations predict the same behavior, indicating that the dynamics only involve the ground and 1st excited state. Between $4\,mV < |V_{dc} \pm V_{ac}| < 6\,mV$, simulations including the 2nd excited state show the same behavior and reproduce the measurement, indicating transitions through the 2nd excited state. At higher voltages ($|V_{dc} \pm V_{ac}| > 6\,mV$), the 3rd excited state also contributes to the dynamics.*

| Parameter | Set 1 | Set 2 |
|---|---|---|
| **Total spin $S$** | 2 | 2 |
| **Spin-polarization tip $\eta$** | 0.514 | 0.86 |
| **Potential scattering $U$** | 0.761 | 0.76 |
| **g-factor** | 2.15 | 2.15 |
| **Magnetic field $B$** | 2 T | 2 T |
| **Uniaxial anisotropy $D$** | -1.692 meV | -1.557 meV |
| **Transversal anisotropy $E$** | 0.262 meV | 0.248 meV |
| **'Sample-sample' conductivity $C_{SS}$** | 7.23 $\mu S$ | 7.2 $\mu S$ |
| **'Sample-tip conductivity $C_{ST}$** | 0.877 $\mu S$ | 0.500 $\mu S$ |
| **Microwave amplitude $V_{ac}$** | 2.8 mV | 1.3 mV |
| **Offset voltage $V_{dc}$** | 4 mV | -- |
| **Temperature** | 2.4 K | 0.5 K |

***Supplementary Table 1 Simulation parameters.*** *The simulations follow references [55] and [56], and are adjusted for a single Fe atom on $Cu_2N$. The displayed parameters are those used to simulate the SRS measurements on the single Fe atom in Fig. 1 of the main text and supplementary Fig. S12. Parameters were obtained by fitting the signal of the numerically evolved dynamics to the measurement in Fig. 1 and adjusting parameters to match the experimental data in supplementary Fig. S12. The linearized conductivity difference is obtained by calculating state-dependent tunneling rates in the relevant voltage range and calculating the average conductivity of each state.*

S1 Solution to the analytic two-state model

A two-state system with Markovian transition rates can be described by the following rate equation:

$$\frac{dn}{dt} = \Gamma_+(t) - \left(\Gamma_+(t) + \Gamma_-(t)\right) \cdot n$$

*[1]*

where $n$ is the occupation probability of one of the states, and $\Gamma_+$ ($\Gamma_-$) the transition rate into (from) that state.

Starting from this rate description, it is convenient to use a rescaled time $\tau = \Gamma t$, with $\Gamma$ being average of the total rate: $\Gamma = \langle \Gamma(t) \rangle = \langle \Gamma_+(t) + \Gamma_-(t) \rangle$. Written with this rescaled time, the rate equation becomes:

$$\frac{dn}{d\tau} = \frac{\Gamma_+(\tau)}{\Gamma} - \frac{\Gamma(\tau)}{\Gamma} n\,.$$

*[2]*

In stochastic resonance spectroscopy (SRS), we modulate the rates with a periodic drive, which makes the rates also periodic with the driving frequency $\omega$. We again use a rescaled frequency $\Omega = \omega/\Gamma$.

The linear, first order differential equation, of eq. [2] with periodic parameters can be solved using [19,20]

$$n(\tau) = \frac{1}{1 - e^{-2\pi/\Omega}} \int_0^{2\pi/\Omega} \frac{\Gamma_+(\tau - \tau')}{\Gamma} e^{-\tau'} h(\tau - \tau', \tau)\, d\tau',$$

*[3]*

$$h(\tau - \tau', \tau) = e^{-\int_{\tau-\tau'}^{\tau} \frac{\Gamma(\tau'')}{\Gamma} - 1 d\tau''}.$$

*[4]*

In the following, we discuss the solution of the differential equation first for the case of constant $\Gamma$, and then we discuss the solution for time-dependent rates $\Gamma(t)$.

S1.1 Calculation of state occupation with constant $\Gamma$

As in the main text, we start with the case where $\Gamma(t) = \Gamma = const$. In this case, the $h$-term in eq. [4] becomes

$$h(\tau - \tau', \tau) = e^{-\int_{\tau-\tau'}^{\tau} 1 - 1 d\tau''} = 1$$

*[5]*

and the occupation probability simplifies to

$$n(\tau) = \frac{1}{1 - e^{-2\pi/\Omega}} \sum_q A_q \int_0^{2\pi/\Omega} e^{iq\Omega(\tau - \tau')} e^{-\tau'}\, d\tau'$$

*[6]*

where the parameters $A_q$ stem from the Fourier series of the rate

$$\frac{\Gamma_+(\tau)}{\Gamma} = \sum_k \frac{a_k}{\Gamma} e^{ik\omega t} = \sum_k A_k e^{ik\Omega\tau}.$$

*[7]*

Note that positive rates imply that $a_k \leq a_0 \leq \Gamma$, i.e. $A_k \leq 1$.

In this form, the integral in eq. *[6* can be evaluated directly, giving:

$$\int_0^{2\pi/\Omega} e^{-(iq\Omega+1)\tau'}\, d\tau' = \frac{1}{-(iq\Omega+1)} \left[e^{-(iq\Omega+1)\tau'}\right]_0^{2\pi/\Omega} = \frac{1}{-(iq\Omega+1)} \left(e^{-(iq\Omega+1)\frac{2\pi}{\Omega}} - 1\right)$$

*[8]*

and thus the occupation reduces to

$$n(\tau) = \sum_q A_q \frac{1}{iq\Omega+1} e^{iq\Omega\tau} = \sum_q \frac{a_q}{\Gamma} \frac{1-iq\Omega}{q^2\Omega^2+1} e^{iq\Omega\Gamma t}.$$

*[9]*

As described in the main text (section III), the experimental signal consists of contributions proportional to the average occupation probability and the in-phase component of the occupation probability, oscillating at the driving frequency. In the case of constant $\Gamma$, the average occupation is the term with $q = 0$ and constant with frequency. The components oscillating at the driving frequency are given by the $q = \pm 1$ terms. The homodyne detection mixes down their in-phase components, i.e. the real part of the $q = \pm 1$ terms. The signal thus becomes

$$I^{\Gamma=const.} = \Delta\sigma V_{ac} \frac{a_{\pm 1}}{\Gamma} \frac{1}{\Omega^2+1} + I_0,$$

*[10]*

where $\Delta\sigma$ is the conductivity difference between the two states, $V_{ac}$ is the amplitude of the voltage modulation and $I_0$ are all frequency-independent contributions to the current. Experimentally, this frequency-independent offset also includes a signal due to rectification of the voltage modulation on a non-linear tunneling characteristic, e.g. caused by the inelastic scattering of the tunneling electrons. Such frequency-independent signals are commonly used to characterize the transmission of the experimental setup [49].

S1.2 Calculation of state occupation with time-dependent $\Gamma(t)$

The derivation in S1.1 outlines the derivation of the occupation probability when $\Gamma(t) = \Gamma = const$. We now consider the general case in which the time-dependence of $\Gamma(t)$ is taken into account.

We start by writing the total rate $\Gamma(t)$ in its Fourier series

$$\frac{\Gamma(\tau)}{\Gamma} = \sum_m \frac{b_m}{\Gamma} e^{im\omega t} = 1 + \sum_{m\neq 0} B_m e^{im\Omega\tau}.$$

*[11]*

Note that positive rates imply that $b_m \leq \Gamma$ (i.e. $B_m \leq 1$), since here $b_0 = \Gamma$.

This enables us to evaluate the integral in the $h$-term, and gives:

$$h(\tau-\tau',\tau) = e^{-\int_{\tau-\tau'}^{\tau}\sum_{m\neq 0}B_m e^{im\Omega\tau''}d\tau''} = e^{-\sum_{m\neq 0}\frac{B_m}{im\Omega}\left(e^{im\Omega\tau}-e^{im\Omega(\tau-\tau')}\right)}. \tag{12}$$

While this is an explicit solution of the $h$-term, the nested exponentials still prevent the straight-forward calculation of the occupation. Thus, we now expand the outer exponential as:

$$h(\tau-\tau',\tau) = \sum_{j=1}\frac{1}{(j-1)!}\left(\frac{1}{i\Omega}\right)^{j-1}\left(-\sum_{m\neq 0}\frac{B_m}{m}\,e^{im\Omega\tau}\left(1-e^{-im\Omega\tau'}\right)\right)^{j-1}. \tag{13}$$

This form shows that, already within the $h$-term, various harmonics of the rate $\Gamma(t)$ mix. This takes place in terms with $j \geq 3$. For instance, the term with order $j=3$ describes mixing between any two harmonics of the drive.

Since we treat real periodic signals, the sum over $m$ converges and can in practice be truncated to a finite order N. This allows us to expand the multinomial in the following way:

$$h(\tau-\tau',\tau) = \sum_{j=1}\frac{1}{(j-1)!}\left(\frac{1}{i\Omega}\right)^{j-1}\sum_{|\boldsymbol{L}|=j-1}\binom{j-1}{\boldsymbol{L}}\prod_{n=1}^{N}\left(\frac{-B_{m_n}}{m_n}\,e^{im_n\Omega}\left(1-e^{-im_n\Omega\tau'}\right)\right)^{L_n}. \tag{14}$$

Here, we adopt a notation with multi-indices $\boldsymbol{L}=(L_1,L_2,\ldots,L_N)$ and $\boldsymbol{m}=(m_1,m_2,\ldots,m_N)$ , where the multinomial coefficients are $\binom{j}{\boldsymbol{L}}=\frac{j!}{L_1!\cdot L_2!\cdot\ldots\cdot L_{j-1}!}$ and $|\boldsymbol{L}|=\sum_{n=1}^{N}L_n$.

Additionally, rearranging the multiplication and introducing the integer $S(\boldsymbol{L},\boldsymbol{m})=\sum_n L_n m_n$ yields

$$h(\tau-\tau',\tau) = \sum_{j=1}\frac{1}{(j-1)!}\sum_{|\boldsymbol{L}|=j-1}\binom{j-1}{\boldsymbol{L}}e^{iS(\boldsymbol{L},\boldsymbol{m})\,\Omega\tau}\left(\frac{1}{i\Omega}\right)^{j-1}\prod_{n}\left(\frac{-B_{m_n}}{m_n}\right)^{L_n}\left(1-e^{-im_n\Omega\tau'}\right)^{L_n} \tag{15}$$

The last step is to expand the terms with $\tau$'

$$h(\tau-\tau',\tau) = \sum_{j=1}\frac{1}{(j-1)!}\sum_{|\boldsymbol{L}|=j-1}\binom{j-1}{\boldsymbol{L}}e^{iS(\boldsymbol{L},\boldsymbol{m})\Omega\tau}\left(\frac{1}{i\Omega}\right)^{j-1}\left(\prod_{n}\left(\frac{-B_{m_n}}{m_n}\right)^{L_n}\right)\sum_{\boldsymbol{l}\leq\boldsymbol{L}}\binom{\boldsymbol{L}}{\boldsymbol{l}}(-1)^{|\boldsymbol{l}|}e^{-iS(\boldsymbol{l},\boldsymbol{m})\Omega\tau'}, \tag{16}$$

which introduces another multi-index $\boldsymbol{l}=(l_1,l_2,\ldots,l_N)$, where the sum over $\boldsymbol{l}\leq\boldsymbol{L}$ covers all $l_n \leq L_n$ and the binomial coefficients are $\binom{\boldsymbol{L}}{\boldsymbol{l}}=\binom{L_1}{l_1}\cdot\binom{L_2}{l_2}\cdot\ldots\cdot\binom{L_N}{l_N}$.

It is now possible to evaluate the integral in eq. [3] by using the Fourier expansion of the excitation rate from eq. [7] and the $h$-term from eq. [16]. The integral over $\tau'$ gives rise to:

$$\int_0^{2\pi/\Omega} e^{-(ik\Omega+1+iS(\boldsymbol{l},\boldsymbol{m})\Omega)\tau'}\,d\tau' = \frac{1}{i(k+S(\boldsymbol{l},\boldsymbol{m}))\Omega+1}\left(1-e^{-(i(k+S(\boldsymbol{l},\boldsymbol{m}))\Omega+1)\frac{2\pi}{\Omega}}\right). \tag{17}$$

As a result, harmonics stemming both from the $h$-term and the excitation rate are mixed giving rise to the following expression for the occupation probability:

$$n(\tau) = \sum_k A_k \sum_{j=1} \frac{1}{(j-1)!} \sum_{|\boldsymbol{L}|=j-1} \binom{j-1}{\boldsymbol{L}} \left( \prod_n \left( \frac{-B_{m_n}}{m_n} \right)^{L_n} \right) e^{i(k+S(\boldsymbol{L},\boldsymbol{m}))\Omega\tau} \left( \frac{1}{i\Omega} \right)^{j-1} \sum_{\boldsymbol{l} \leq \boldsymbol{L}} \binom{\boldsymbol{L}}{\boldsymbol{l}} \frac{(-1)^{|\boldsymbol{l}|}}{i(k+S(\boldsymbol{l},\boldsymbol{m}))\Omega + 1}$$

$$= \sum_k \sum_{j=1} \sum_{|\boldsymbol{L}|=j-1} \binom{j-1}{\boldsymbol{L}} \frac{a_k}{\Gamma} \prod_n \left( \frac{-b_{m_n}}{\Gamma} \right)^{L_n} e^{i(k+S(\boldsymbol{L},\boldsymbol{m}))\Omega\Gamma t} \frac{1}{(j-1)!\, i^{j-1} \Omega^{j-1} \prod_n m_n^{L_n}} \sum_{\boldsymbol{l} \leq \boldsymbol{L}} \binom{\boldsymbol{L}}{\boldsymbol{l}} \frac{(-1)^{|\boldsymbol{l}|}}{i(k+S(\boldsymbol{l},\boldsymbol{m}))\Omega + 1}$$

*[18]*

This expression is rather complex as it describes the very general case of arbitrary $\Gamma(t)$. In practice, the rates are often dominated by a finite order in the mixing of harmonics of the drive and all harmonics are smaller or equal to the average of the rate ($b_m, a_k \leq \Gamma$), i.e. the series converges. It is therefore meaningful to treat the sum over $j$ that represents the number of mixing Fourier components as a series expansion of the occupation and to neglect high orders of $j$.

Explicitly, the terms up to $j = 2$ are:

$$n(t) = \underbrace{\sum_k \frac{a_k}{\Gamma} \frac{1 - ik\Omega}{k^2\Omega^2 + 1} e^{ik\Omega\Gamma t}}_{j=1} + \underbrace{\sum_k \sum_{m \neq 0} \frac{a_k}{\Gamma} \frac{-b_m}{\Gamma} \frac{1}{m} \left[ \frac{(m+k) + \frac{i}{\Omega}}{(m+k)^2\Omega^2 + 1} - \frac{k + i/\Omega}{k^2\Omega^2 + 1} \right] e^{i(k+m)\Omega\Gamma t}}_{j=2}$$

$$+ \underbrace{O\left( \frac{a_k}{\Gamma} \left( \frac{b_m}{\Gamma} \right)^2 \right)}_{j>2}$$

*[19]*

## S2 Occupation in the high and low frequency limit

A physical understanding of the dynamics at play can be gained by looking at the occupation probability in the high and low frequency limits (see also Fig. 2 of the main text).

### S2.1 Occupation at $\Omega \gg 1$

At high frequencies, we have $\frac{1}{\Omega^{j-1}} \to 0$ for $j > 1$. The only term remaining in eq. [18] is thus the one with $j = 1$ (i.e. the first sum in eq. [19]). As a result, Fourier components with $k \neq 0$ oscillate at $k\Omega$ averaging to 0 and the only component in the occupation remaining is that with $k = 0$ leading to:

$$n^{\Omega\gg 1} = \frac{a_0}{\Gamma} = \frac{\langle \Gamma_+(t) \rangle}{\langle \Gamma(t) \rangle}. \tag{20}$$

At high frequencies, the occupation probability thus only depends on the average of the rates: their modulation over time is too fast for the system to react.

### S2.2 Occupation at $\Omega \ll 1$

For small frequencies, we find numerically that the terms in eq. [18] also simplify. More precisely, we verified that up to $j = 13$ (covering most physical scenarios) the following term:

$$\frac{1}{(j-1)!\, i^{j-1} \Omega^{j-1} \prod_n m_n^{L_n}} \sum_{\boldsymbol{l} \leq \boldsymbol{L}} \binom{\boldsymbol{L}}{\boldsymbol{l}} \frac{(-1)^{|\boldsymbol{l}|}}{i\big(k + S(\boldsymbol{l}, \boldsymbol{m})\big)\Omega + 1} \tag{21}$$

tends to unity. In the case of $j = 2$, this is shown in supplementary Fig. S6: there we display the first 10 $k$-terms of the $j = 2$ sum of eq. [18]. Assuming that this finding applies for all $j$, the occupation becomes

$$n^{\Omega\ll 1}(\tau) = \sum_k \frac{a_k}{\Gamma} e^{ik\Omega\tau} \sum_{j=1} \sum_{|\boldsymbol{L}|=j} \binom{j-1}{\boldsymbol{L}} \prod_n \big(-B_{m_n}\big)^{L_n} e^{iS(\boldsymbol{L},\boldsymbol{m})\Omega\tau}, \tag{22}$$

where we can identify the multinomial expansion and get

$$n^{\Omega\ll 1}(\tau) = \sum_k \frac{a_k}{\Gamma} e^{ik\Omega\tau} \sum_{j=1} \left( \sum_{m\neq 0} -B_m e^{im\Omega\tau} \right)^{j-1}. \tag{23}$$

Here, the sum over $j$ is the geometric series of the Fourier series of the rate and we have

$$n^{\Omega\ll 1}(\tau) = \frac{\sum_k \frac{a_k}{\Gamma} e^{ik\Omega\tau}}{1 - \sum_{m\neq 0} -b_m/\Gamma e^{im\Omega\tau}} = \frac{\Gamma_+(t)}{\Gamma(t)}. \tag{24}$$

The occupation is therefore the ratio of the excitation rate $\Gamma_+(t)$ to the total rate $\Gamma(t)$. This result is identical with the steady state solution of the rate equation:

$$\frac{dn}{dt} = 0 = \Gamma_+(t) - \Gamma(t)n. \tag{25}$$

S3 Numerical simulation of dynamics

An alternative approach to obtain the SRS signal theoretically is to simulate the evolution of the occupation probability numerically. In order to do this, we start by mapping the voltage dependent transition rates onto the oscillation period of the drive voltage.

Using these time dependent rates, we simulate the time-dependent occupation probability for any microwave frequencies $\omega$ by evolving the rate equation describing the dynamics (two-state system: eq. [1], multi-state system: eq. [26]) starting from an arbitrary initial occupation distribution. To remove any dependence on the initial occupation distribution, we let the system evolve until the norm of the difference between the occupation of one state during two consecutive periods is smaller than a tolerance value set between $10^{-5}$ and $10^{-7}$.

To speed up simulation time, we first simulate the rate equation for the lowest microwave frequency, where the system stays closest to its steady state solution (see S2.2): and choose therefore the steady-state distribution without drive as our initial state. For each consecutive frequency we then use the occupation of the closest, already simulated, frequency as the initial state.

The time dependent current is calculated from the evolution of the state occupations over the next 3 to 20 periods. When simulating the two-state system we assume generic state-dependent conductivities and thus calculate the current $I = (V_{dc} + V_{ac} \cdot \cos(2\pi f t))(\sigma_g + \Delta\sigma n(t))$ as introduced in the main text. When simulating the multi-state system, we obtain both the transition rates as well as the state-dependent current directly from scattering theory. In the following section, we discuss this case in more detail.

## S4 Numerical simulation of multi-state systems

To derive our analytical solution to the SRS signal, we model the underlying dynamics using a two-level system and rely on the fact, that the dynamics of many multi-level systems can be reduced to effective rates between only two levels and including the influence of all other states by allowing multi-state transitions. To assess how well this approximation fares for the dynamics of the Fe atom as shown in Fig. 1 and 4 of the main text, we also simulate the dynamics of the full 5-state system numerically.

We consider voltage-dependent transition rates $\Gamma_{fi}(V)$ from any initial state $i$ to any final state $f$ of a n-state system: $\Gamma_{fi}(t,\omega) = \Gamma_{fi}(V_{dc} + V_{ac}\cos(\omega t))$ and calculate the voltage-dependent scattering rates according to the scattering theory described in references [55] and [56].

We use these time dependent rates, to simulate the time-dependent occupation probabilities for any microwave frequencies $\omega$ by evolving the rate equation

$$\begin{pmatrix} \frac{dn_1}{dt} \\ \vdots \\ \frac{dn_n}{dt} \end{pmatrix} = \begin{pmatrix} -\sum_{i=2}^{n}\Gamma_{i1}(t,\omega) & \Gamma_{1\,2}(t,\omega) & \cdots & \Gamma_{n\,1}(t,\omega) \\ \Gamma_{2\,1}(t,\omega) & \ddots & \ddots & \vdots \\ \vdots & \ddots & \ddots & \Gamma_{n-1\,n}(t,\omega) \\ \Gamma_{n\,1}(t,\omega) & \cdots & \Gamma_{n\,n-1}(t,\omega) & -\sum_{i=1}^{n-1}\Gamma_{in}(t,\omega) \end{pmatrix} \begin{pmatrix} n_1 \\ \vdots \\ n_n \end{pmatrix}. \quad [26]$$

Similar to the transition rates, we calculate the state-dependent current $I_i$ according to scattering theory and obtain the current with $I(t,\omega) = \sum_{i=1}^{n} I_i(V_{dc} + V_{ac} \cdot \cos(\omega\, t))\, n_i(t,\omega)$.

Using the parameters listed in table 1 we can quantitatively reproduce the SRS measurement of the single Fe atom in Fig. 1 (Parameter set 1) as well as the offset voltage dependence in Fig. 4 (Parameter set 2) of the main text. The results are displayed in supplementary Fig. S11 and S12.

In addition to the simulation with all 5 states, supplementary Fig. S11 also displays simulations, where we successively exclude the state with highest energy.

Comparing the full five-state model to the reduced two-level description shows that signal amplitudes of the two-level model are up to 10% larger than those of the full simulation, but the characteristic frequency of the resulting SRS steps, i.e. the Fe atom's average transition rates, agree within <2% with the full five-state calculation. The rate obtained by fitting the full simulation with eq. 1 of the main text is $1.77 \cdot 10^9/s$, and the rate of the effective two state system is $1.74 \cdot 10^9/s$. This highlights that straightforward fitting of SRS signals with eq. 1 of the main text yields accurate rates even for multi-state spin systems.

The five-state simulation also fits the measured variation of the SRS signal with drive offset voltage and reproduces the peak-dip structures of both thresholds quantitatively (supplementary Fig. S12a, green curve). This enables the identification of the spin states that participate in the magnetization reversal of the Fe atom. Simulating only the ground and first excited state creates a weak SRS Signal and reproduces the peak-dip feature at $\pm 2\ mV$ (yellow curve) indicating that it results from direct transitions between the two low-energy spin states of the easy-axis spin system. Beyond $\pm 3\ mV$, we need to include the second excited state, which lies $4\ mV$ above the ground state. This increases the SRS amplitude making it comparable to the measured one (red curve). It also creates the peak in the SRS signal amplitude at $-5\ mV$, and corresponds to excitations over the anisotropy barrier via higher excited spin states. The third excited state at $6\ mV$ above the ground state is required to correctly reproduce the signal decay to higher offset voltages (blue curve).

We note that the rate shown in supplementary Fig. S12b displays a sharp feature at $+5.0\ mV$, which we suspect occurs when two transition pathways have the same rate and block each other. While the small SRS signal amplitudes in that bias region do not allow us to definitively resolve this feature, it opens up an interesting pathway for future SRS investigations of potential destructive interferences between transport channels.

## S5 Methods

### S5.1 Experimental setup

The measurements on the spin systems, i.e. individual Fe atoms and the 4-atom antiferromagnet, were performed using a low temperature, ultra-high vacuum scanning tunneling microscope (Unisoku USM-1300) with a $^3$He cryostat, a (9/2/2) T superconducting vector magnet and high frequency wiring to the (PtIr)-tip. The measurements on the Yu-Shiba-Rusinov (YSR) state were performed using an ultra-low temperature, ultra-high vacuum scanning tunneling microscope (Unisoku USM-1600) with a dilution refrigerator and high frequency wiring to a bulk Vanadium-tip.
In all measurements, the low-frequency offset voltage was combined with the high-frequency microwave modulation using a bias-tee (Spin and Néel state measurements: PSPL5575A-104, YSR measurements: Anritsu V251) and was applied to the tip-side of the tunnel junction. The frequency-dependent microwave modulation was produced by an arbitrary waveform generator (Keysight M8195m).

### S5.2 Sample preparation

The sample surfaces were cleaned by repeated cycles of $Ar^+$ sputtering at 1 kV at a background pressure of $2 \cdot 10^{-6}$ mBar and subsequent annealing to several hundred K, respectively. The Cu(100) surface, for the measurements on the spin systems, was annealed at 850 K. The thin $Cu_2N$ decoupling layer was grown by sputtering the surface at 1 kV in a $N_2$ atmosphere of $5 \cdot 10^{-6}$ mBar and consecutive annealing at 600 K. The V(100) surface, for the measurements on the YSR state, was annealed at 1073 K. For both substrates, Fe atoms were deposited onto the prepared and precooled sample surfaces (~20 K). The sample was positioned in front of a Knudsen cell evaporator with metallic Fe heated to 1300 K, for 8 s for the $Cu_2N$ sample and 1500 K for 20 s for the V(100) sample.

### S5.3 Atom manipulation

On the $Cu_2N$ decoupling layer, Fe atoms occur naturally as individual atoms and can be identified by their spectroscopic features using conductance measurements [40]. Chains, consisting of multiple Fe atoms were assembled by picking up and placing Fe atoms via vertical manipulation. The Fe atoms in the chain are arranged 0.72 nm apart and parallel to the external magnetic field [8].